\documentclass[12pt]{article} 
\usepackage{graphicx} 
\makeatletter 
\newenvironment{mybibliography}[1]{
\section*{References}\list{\@biblabel{\@arabic\c@enumiv}}%
           {\settowidth\labelwidth{\@biblabel{#1}}%
            \leftmargin\labelwidth 
            \advance\leftmargin\labelsep 
            \@openbib@code 
            \usecounter{enumiv}%
            \let\p@enumiv\@empty 
            \renewcommand\theenumiv{\@arabic\c@enumiv}}%
    \sloppy 
      \clubpenalty4000 
      \@clubpenalty \clubpenalty 
      \widowpenalty4000%
      \sfcode`\.\@m} 
     {\def\@noitemerr 
       {\@latex@warning{Empty `thebibliography' environment}}%
      \endlist} 
\def\ps@myheadings{%
    \let\@oddfoot\@empty\let\@evenfoot\@empty
    \def\@evenhead{\thepage\hfil\slshape\leftmark}%
    \def\@oddhead{{\slshape\rightmark}\hfil\mbox{}}%
    \let\@mkboth\@gobbletwo
    \let\sectionmark\@gobble
    \let\subsectionmark\@gobble
    }
\makeatother 

\newcommand{\mymatrix}[1]{\widetilde{#1}} 
\newcommand{\vecm}{{\rm\bf m}}
\newcommand{\mathi}{{\rm i}}  
\newcommand{\veci}{{\rm\bf i}}  
\newcommand{\vecr}{{\rm\bf r}}  
\newcommand{\vecj}{{\rm\bf j}}  
\newcommand{\vecn}{{\rm\bf n}}  
\newcommand{\veck}{{\rm\bf k}}  
\newcommand{\vecp}{{\rm\bf p}} 
\newcommand{\vech}{{\rm\bf h}} 
\newcommand{\vecc}{{\rm\bf c}} 
  
\newcommand{\fsgn}{{\rm fsgn}}  
\newcommand{\sumprime}{\mathop{\sum{}'}}  
\newcommand{\prodprime}{\mathop{\prod{}'}}

\begin{document} 
\begin{center} 
\mbox{}\\[6cm] 
{\large\bf GUTZWILLER-CORRELATED WAVE FUNCTIONS:}\\[6pt] 
{\large\bf APPLICATION TO FERROMAGNETIC NICKEL}\\[2\baselineskip] 
J\"org B\"unemann$^1$, Florian Gebhard$^1$, Torsten Ohm$^2$,  
Stefan Weiser$^2$, and Werner Weber$^2$\\[\baselineskip] 
$^1$ Fachbereich Physik and Material Sciences Center,\\ 
Philipps-Universit\"at Marburg, D-35032 Marburg,  
Germany\\[\baselineskip]  
$^2$ Institut f\"ur Physik, \\Universit\"at Dortmund, D-44221 Dortmund,  
Germany\\[2cm] 
\end{center} 

\section{Introduction}
\label{Sec:Introduction}

\subsection{Ferromagnetism: whither theory?}
\label{subsec:whither}

Metallic ferromagnetism has since long been a controversial subject in
solid-state theory. From early on, two schools of thought have been
competing. The first, based on ideas of Slater, Stoner, and Wohlfarth,
turns to the band aspects of metallic ferromagnets such as iron, cobalt,
and nickel~\cite{Wohlfarth}. 
The itineracy of the conduction electrons, including those
of the open 3$d$~shell, is considered to dominate the
electron-electron interaction. The second school, going back
to van-Vleck~\cite{vanVleck} and Gutzwiller~\cite{Gutzwiller1963}, 
stresses the localized
character of the 3$d$~electrons in transition metals and their compounds; 
local magnetic moments are indeed manifest in magnetic insulators
which are often oxides and halides of transition metals.
In the van-Vleck--Gutzwiller school,
local moments are seen as an important source of magnetism,
common to both ferromagnetic transition metals
and their insulating compounds.
Therefore, the van-Vleck--Gutzwiller school emphasizes
the importance of correlations
as a consequence of the electron-electron interactions.

As the Slater--Stoner--Wohlfarth school 
considers electronic correlations to be of minor importance,
it treats the electron-electron interaction in simple
approximations such as Hartree--Fock theory, see~\cite{Wohlfarth}.
Later, the density-functional theory (DFT) in its spin-dependent versions
(SDFT) took over the role of Hartree--Fock or $x_{\alpha}$~schemes
of the Slater school. For more than two decades, SDFT energy-band
calculations by various groups have provided reasonably accurate results
for quite a number of properties of metallic ferromagnets, 
e.g., magnetic moments and the overall shape of the Fermi surface.
An early problem was the observation that SDFT in standard
local-density approximation (LDA) does not give the bcc~lattice
as the most stable crystal structure for iron, instead fcc~Fe
is found to be stable. This problem was overcome by the introduction
of generalized gradient expansions (GGA) of the density~\cite{Perdew}.

Until recently
the van-Vleck--Gutzwiller school could not provide
much more than intuitive arguments for its case because ferromagnetism
is seen as a genuine, difficult many-body problem.
For decades, only simplistic models could be treated with
some reliability, e.g., the one-band Hubbard model,
which, however, exhibits ferromagnetism only for an extremely
large Hubbard interaction or a pathological density of 
state~\cite{Volloneband}. 
In the past, despite some partial successes~\cite{Nolting}, 
correlated-electron theory could not describe real materials
in much detail. Over the last decade new theoretical tools have been
developed which try to combine the merits of density-functional theory
with those of advanced many-body approaches, e.g., the LDA-$GW$ 
approximation~\cite{GW1}.
A general problem of correlated-electron theories
is the construction of suitable model Hamiltonians.
Progress in this direction has been made
recently, e.g., the `down-folding' scheme 
permits a systematic construction of 
Hamiltonians in a reduced Hilbert space~\cite{Andersen}.
The one-particle dynamics of models for correlated lattice electrons 
can be determined, in principle, within
the dynamical mean-field theory (DMFT)~\cite{DMFT,Zoelfl,Lichtenstein}
which becomes exact in the limit of an infinite number of nearest neighbors
(coordination number $Z\to\infty$). 
The `static approximation' to DMFT is the `LDA+$U$' scheme 
which has been applied to a variety of correlated-electron 
systems~\cite{LDApU,KotliarLDApU,Xie}.
Moreover, DMFT has been extended
in various ways, e.g., a combination of
the LDA-$GW$ approximation and the DMFT has
been put forward~\cite{GW2}.
Therefore, correlated-electron theories
seem to approach a level of quality where a faithful comparison 
with experiments becomes possible.

In our group, we have extended the original
Gutzwiller variational scheme to the full multi-band problem.
Our theory provides total energies and
magnetic moments of competing magnetic and non-magnetic phases
and the single-particle excitation energies within 
Fermi-liquid theory. Thereby, our theory describes the dispersion of
the conduction and valence bands and the Fermi surface as accessible,
e.g., by angle-resolved photo-emission spectroscopy (ARPES),
and the predictions of our Gutzwiller correlated-electron theory 
can be tested against experiments on transition metals and its compounds.

\subsection{The nickel problem}
\label{subsec:nickelproblem}

In this article we address ferromagnetic nickel as the prototype
of band ferromagnetic materials. On the experimental side,
one of the reasons why nickel
has been in the focus of interest, much more than iron or cobalt,
is the relative ease to grow large single crystals. 
Consequently,
a lot of experimental investigations have been carried out on nickel
in order to determine its magnetic properties and electronic structure. 
For a review, we refer to the 
articles~\cite{Wohlfarth,Donath,LB19,LB23}.

On the theoretical side, of all the iron group magnetic metals
nickel is the most celebrated case of discrepancies between SDFT
predictions and experimental results. 
Early on, de-Haas--van-Alphen
data indicated the presence of only one hole ellipsoid in the Fermi
surface of the minority spin bands, located around the $X$-point of
the Brillouin zone~\cite{Tsui}. In contrast, energy band calculations
using SDFT predict two hole ellipsoids around
the $X$-point, the $X_5$-state of pure $d(t_{2g})$ orbital character,
and the $X_2$-state of pure $d(e_g)$ orbital character.
Concomitantly, neutron scattering experiments~\cite{Mook} 
revealed that the orbital
character of the nickel magnetic moment exhibits
less $e_g$ admixture than predicted by energy band
calculations.

ARPES measurements confirmed
the de-Haas--van-Alphen results: the minority $X_2$-state of pure 
$d(e_g)$ orbital character was found to lie slightly below the Fermi
level. Furthermore, the photoemission data differ
considerably from SDFT over the whole range of occupied $3d$~bands.
Most importantly, the width of the occupied part of the 3$d$ bands
is approximately $W=3.3\, {\rm eV}$~\cite{EHK78,EP80}, 
whereas SDFT gives $W_{\rm SDFT}=4.5\, {\rm eV}$, or larger~\cite{Moruzzi}.
ARPES studies also reveal big discrepancies between SDFT and
experiment in the exchange splittings of majority and minority spin
bands near the Fermi energy. The SDFT results give a rather isotropic
exchange splitting of about $650\, {\rm meV}$ to $750\, {\rm meV}$.
whereas photoemission data show small and highly anisotropic
exchange splittings between $160\, {\rm meV}$ for pure $3d(e_g)$-states
such as $X_2$ and $330\, {\rm meV}$ for pure $3d(t_{2g})$-states
such as $S_3$ near the $X$-point. 

Another discrepancy concerns the position of the $L_{2'}$-state.
Experimentally, it is found approximately~$1\,{\rm eV}$ below
the Fermi energy~$E_{\rm F}$ whereas the SDFT puts it to 
about~$0.3\, {\rm eV}$ below~$E_{\rm F}$. This state is special as it
represents the most binding, pure~$4p$-state~\cite{4dTM}. 
An incorrect $4p$-level implies that DFT slightly underestimates the
partial density of the $4p$~electrons and, thus, the $3d$-hole density.

Lastly, the size and the direction of the magnetic moment
pose a big problem to SDFT.
Early experiments using ferromagnetic electron resonance and the
Einstein--de-Haas effect have produced the values $g=2.183$ and
$g'=1.835$ which result in an orbital moment of
$\mu_{\rm orb}=0.0507 \mu_{\rm B}$ per atom for nickel~\cite{Wohlfarth}. 
In general, the values of the orbital part of the magnetic moments in band
ferromagnets are not very well reproduced in SDFT.
Iron and cobalt show large deviations~\cite{Singh}, even in relativistic
SDFT calculations~\cite{Dalderop}. 
Moreover, the magneto-crystalline anisotropy energies
are not correct, i.e., they come with the wrong sign in SDFT.
As shown by Aubert and Gersdorf~\cite{Gersdorf,Gersdorf2},
the value of the dominant anisotropy constant is~$K_1=-8.8\, \mu{\rm eV}$,
i.e., the (111)~direction is the easy axis in nickel.
For a (001) magnetic moment, the energy is higher with an energy difference
$E_{\rm aniso}\equiv E_{111}-E_{001}\approx -3\, \mu{\rm eV}$ per atom.
In contrast, it is $E_{\rm aniso}>0$ in SDFT calculations, e.g., by
Daalderoop~\cite{Dalderop} so that $K_1>0$.
The SDFT results for cobalt and iron are equally disappointing.
At low temperatures, the complex behavior of the magnetic anisotropy 
as a function of temperature and the magnetic field was 
interpreted by Gersdorf~\cite{Gersdorf} to be caused by
a shift of the electronic state (001)~$X_{2\downarrow}$
to about $3\,{\rm meV}$ {\sl above\/} the Fermi energy.
The famous second hole ellipsoid around the~$X_{2\downarrow}$
indeed exists~\cite{Gersdorf} as long as the magnetic moment
has an angle of less than 18~degrees with the (001)~direction.
However, the states 
(100)~$X_{2\downarrow}$ and (010)~$X_{2\downarrow}$ remain
below~$E_{\rm F}$, about $33\, {\rm meV}$~\cite{Gersdorf}.

All these discrepancies could not be overcome
by SDFT modifications such as the GGA variants~\cite{Perdew}.
Apparently, a substantial improvement of SDFT results for nickel
along these lines is presently not within reach.
Recent correlated-electron theories remove the 
SDFT shortcomings only partially. For example, 
with LDA+$U$-type calculations it seems to be possible to
adjust to experiment the orbital moment, the magnetic
anisotropy, and the Fermi surface topology~\cite{KotliarLDApU,Xie};
however, other quantities, especially the entire quasi-particle band structure,
remain elusive within these correlated-electron theories.

\subsection{Scope of this work}
\label{subsec:plan}

In this article we show that our Gutzwiller theory resolves all
of the discrepancies described in Section~\ref{subsec:nickelproblem}. 
Gutzwiller theory reproduces the experimental
Fermi surface topology, the quasi-particle band structure with 
the correct width and the anisotropic exchange splittings, and
the magnetic anisotropy. This success for nickel corroborates
the early views by van-Vleck and Gutzwiller:
the inclusion of the electron-electron interaction 
on a many-body level is essential for a proper understanding
of itinerant ferromagnetism.

We structure our article as follows. In Section~\ref{GWF}
we introduce our general Gutzwiller-correlated wave functions for
multi-band Hubbard models. In Section~\ref{results} 
we present analytical results in the
limit of infinite coordination number which we apply
to nickel.
The remaining Sections~\ref{GWFevaluate}--\ref{LandauGutzwiller}
are devoted to an improved, concise derivation of our theory.
A short outlook closes our presentation.

The class of variational wave functions for which our formulae apply
goes much beyond our previous
investigations~\cite{PRB,EPL,thulpaper}.
First, the Gutzwiller
correlator now employs projections onto eigenstates of
an effective atomic Hamiltonian which makes the variational
Hilbert space much more flexible with respect to local correlations.
Second, our variational states cover the case of
superconductivity in correlated multi-band lattice systems. 
For ferromagnetic nickel, it does not seem to be necessary to exploit fully 
these new flexibilities. Therefore, we restrain ourselves to
the normal-conducting state.

\section{Gutzwiller variational theory}  
\label{GWF}

In this section we first define Gutzwiller variational wave functions
for multi-band Hubbard models.
In the limit of infinite coordination number, $Z\to\infty$,
the expectation value
of the Hamiltonian can be written in terms of an
effective single-particle Hamiltonian. We use this expression
as an approximation in our numerical studies on fcc nickel ($Z=12$). 
 
\subsection{Definitions} 
\label{subsec:defs}
 
\subsubsection{Gutzwiller-correlated wave functions}

We consider the Gutzwiller-correlated wave functions  
\begin{equation}  
| \Psi_{\rm G}\rangle \equiv \hat{P}_{\rm G} | \Psi_0\rangle \; .  
\label{defwavefunction}  
\end{equation}  
Here, $| \Psi_0\rangle$ is a normalized
quasi-particle vacuum so that we may  
later apply Wick's theorem. Expectation values in $| \Psi_0\rangle$  
are denoted by  
\begin{equation}  
\langle \hat{A}\rangle_0 \equiv \langle \Psi_0 | \hat{A}| \Psi_0\rangle  
 \; .  
\end{equation}  
Expectation values in $| \Psi_{\rm G}\rangle$  
are denoted by  
\begin{equation}  
\langle \hat{A}\rangle_{\rm G}   \equiv
\frac{\langle \Psi_{\rm G} | \hat{A} |\Psi_{\rm G}\rangle}{  
\langle \Psi_{\rm G} | \Psi_{\rm G}\rangle}  
 \; .  
\end{equation}  
The Gutzwiller correlator is a product of local correlators,   
\begin{equation}  
 \hat{P}_{\rm G}  \equiv \prod_{\vecm}  \hat{P}_{\rm G|\vecm} \; ,  
\end{equation}  
which induce transitions between atomic configurations on lattice  
site~$\vecm$,  
\begin{equation}  
\hat{P}_{\rm G|\vecm} \equiv \sum_{I_1,I_2} \lambda_{I_1,I_2|\vecm}  
|I_1\rangle_{\vecm}   
{}_{\vecm}\langle I_2|  \; .  
\label{defPGl}  
\end{equation}  
Atomic configurations with at least one electron are defined by  
\begin{equation}  
|I\rangle_{\vecm} \equiv \prod_{\sigma\in I} \hat{c}_{\vecm,\sigma}^+  
|\hbox{vacuum}\rangle_{\vecm} \; .  
\end{equation}  
The operator $\hat{c}_{\vecm,\sigma}^+$ ($\hat{c}_{\vecm,\sigma}$) 
creates (annihilates) an electron on lattice site~$\vecm$ in the 
local spin-orbitals $\sigma=1,2,\ldots,(2N_{\rm o})$ 
($N_{\rm o}=1,3,5$ for $s,p,d$~electrons). 
The sets~$I$ comprise all possible atomic occupancies with  
$|I|=1,2,\ldots, (2N_{\rm o})$ electrons.   
$I=(\emptyset),(\sigma_1), (\sigma_1,\sigma_2),
\ldots,$ $(\sigma_1,\ldots,\sigma_{2N_{\rm o}})$ 
denotes configurations with no, one, two, \ldots, $2N_{\rm o}$ electrons. 
A fixed order among the orbitals is assumed, i.e., we have
$\sigma_1<\sigma_2<\ldots<\sigma_{2N_{\rm o}}$ in the definition of the sets
 $I$. For example, there is only one configuration 
$(2N_{\rm o})\equiv(\sigma_1,\ldots,\sigma_{2N_{\rm o}})=
(1,\ldots,2N_{\rm o})$ 
which represents a fully occupied atom. 
All standard set operations are defined for the sets $I$, e.g.,
$I=I_1\cup I_2$ is the set union, and $I=I_1\setminus I_2$
denotes the set of elements in $I_1$ which are not in $I_2$.
We define the complement of $I$ as 
$\bar{I}\equiv (2N_{\rm o})\setminus I$. Altogether there are 
$D_{\rm at}=2^{2N_{\rm o}}$ atomic configurations, 
$D_{\rm at}=4,64,1024$ for $s,p,d$~electrons.  
  
We demand that $|I_1|+|I_2|$ is even in~(\ref{defPGl}).  
This restriction permits pairing correlations with an   
even number of Fermions as in BCS theory.   
Furthermore, we address only Hermitian Gutzwiller correlators,  
$\hat{P}_{\rm G|\vecm} =   
\hat{P}_{\rm G|\vecm}^+$. Consequently, we demand  
\begin{equation}  
 \lambda_{I_1,I_2|\vecm}
=  \lambda_{I_2,I_1|\vecm}^*\; .  
\end{equation}  
Therefore, we can find atomic states $|\Gamma\rangle_{\vecm}$  
such that  
\begin{equation}  
\hat{P}_{\rm G|\vecm} = \sum_{\Gamma} \lambda_{\Gamma|\vecm}
\hat{m}_{\Gamma|\vecm}
\end{equation}  
with real parameters $ \lambda_{\Gamma|\vecm }$.
Here, 
\begin{equation}  
\hat{m}_{\Gamma|\vecm} = |\Gamma\rangle_{\vecm}   
{}_{\vecm}\langle \Gamma|   
\end{equation}  
is the projection operator onto the state 
$|\Gamma\rangle_{\vecm}$.
In terms of atomic configurations we have 
\begin{equation}  
|\Gamma\rangle_{\vecm} = \sum_I A_{I,\Gamma|\vecm}|I\rangle_{\vecm}  
\quad , \quad 
|I\rangle_{\vecm} = \sum_{\Gamma} A_{I,\Gamma|\vecm}^*|\Gamma\rangle_{\vecm}  
\end{equation}  
where the unitary matrix $\mymatrix{A}_{\vecm}$ diagonalizes the matrix   
$\mymatrix{\lambda}_{\vecm}$,  
\begin{equation}  
\lambda_{\Gamma_1|\vecm} 
\delta_{\Gamma_1,\Gamma_2}  =
\sum_{I_1,I_2}A_{I_1,\Gamma_1|\vecm}^*  \lambda_{I_1,I_2|\vecm}  
 A_{I_2,\Gamma_2|\vecm} \;.  
\end{equation}  
In general, the state $|\Gamma \rangle_{\vecm}$ can be chosen
as the eigenstates of some {\sl effective\/} local atomic 
Hamiltonian~$\hat{H}_{\rm at}^{\rm eff}$.

\subsubsection{Multi-band Hubbard models}

In this work we focus on ferromagnetism of transition metals.  
Therefore, we study the multi-band Hubbard model for $N$~electrons
on~$L$ lattice sites,  
\begin{equation}  \label{hamiltonian5}
\hat{H}= \hat{T} +\hat{H}_{\rm at}  
\end{equation}  
with  
\begin{eqnarray}  
\hat{T}&=& \sum_{\veci,\vecj,\sigma \sigma'} 
 t_{\veci,\vecj}^{\sigma \sigma'}   
\hat{c}_{\veci,\sigma}^+\hat{c}_{\vecj,\sigma'} 
=
\sum_{\veck;\sigma \sigma'}\epsilon_{\sigma \sigma'}^0(\veck)  
\hat{c}_{\veck,\sigma}^+\hat{c}_{\veck,\sigma'}\;,\\ 
\epsilon_{\sigma \sigma'}^0(\veck)&\equiv&
\frac{1}{L}\sum_{\veci,\vecj}e^{-\mathi\veck(\vecj-\veci)}
t_{\veci,\vecj}^{\sigma \sigma'} \; ,
\label{bareen}
\end{eqnarray}  
where we used translational invariance, 
$t_{\veci,\vecj}^{\sigma \sigma'} \equiv t^{\sigma \sigma'}(\veci-\vecj)  $
to introduce the bare energy-band matrix 
elements~$\epsilon_{\sigma \sigma'}^0(\veck)$. Note that 
we set $ t_{\veci,\veci}^{\sigma \sigma'} =0$, i.e., the kinetic energy
operator does not include the local potential terms
$\epsilon_{\veci,\sigma\sigma'}\equiv \epsilon_{\sigma\sigma'}$
(e.g., exchange splittings, spin-orbit interaction)
which are contained in the general, purely local Hamiltonian
\begin{equation}  
\hat{H}_{\rm at} = \sum_{\vecm} \sum_{I_1,I_2} U_{I_1,I_2|\vecm}  
|I_1\rangle_{\vecm}   
{}_{\vecm}\langle I_2| =
\sum_{\vecm} \sum_{\overline{\Gamma}}E_{\overline{\Gamma}|\vecm}
\hat{m}_{\overline{\Gamma}|\vecm} \; .
\label{Hlocalnew}  
\end{equation}  
The projector onto the atomic eigenstate~$|\overline{\Gamma}\rangle_{\vecm}$
on site~$\vecm$ is denoted $\hat{m}_{\overline{\Gamma}|\vecm}$.
The unitary matrix $\mymatrix{B}_{\vecm}$ 
diagonalizes the matrix $\mymatrix{U}_{\vecm}$,
\begin{equation}  
E_{\overline{\Gamma}_1|\vecm} 
\delta_{\overline{\Gamma}_1,\overline{\Gamma}_2}
= \sum_{I_1,I_2} B_{I_1,\overline{\Gamma}_1|\vecm}^*
U_{I_1,I_2|\vecm} B_{I_2,\overline{\Gamma}_2|\vecm} \;  .
\end{equation}
For translation invariance, we have $E_{\overline{\Gamma}|\vecm} 
\equiv E_{\overline{\Gamma}}$,
$U_{I_1,I_2|\vecm}  \equiv U_{I_1,I_2}$,
$ \lambda_{I_1,I_2|\vecm}\equiv  \lambda_{I_1,I_2}$, and
$\lambda_{\Gamma|\vecm} \equiv  \lambda_{\Gamma}$.
Note that the eigenstates~$|\overline{\Gamma}\rangle_{\vecm}$
of $\hat{H}_{\rm at}$ need not be identical to 
$|\Gamma\rangle_{\vecm}$.
Therefore, the number of parameters $\lambda_{I_1,I_2}$  
in our general definition of the Gutzwiller correlator~(\ref{defPGl}) 
is of the order of $D_{\rm at}^2$ [$={\cal O}(10^6)$ for $d$~electrons].  
In most practical applications, it is impossible to scan such 
a large variational space numerically. Therefore, one has to work 
with eigenstates $|\Gamma\rangle_{\vecm}$  
of some {\em effective\/} atomic Hamiltonian 
$\hat{H}_{\rm at}^{\rm eff}$ which is characterized by a few 
physically motivated parameters.  
In this way, the numerical problem basically reduces  
to the minimization of the ground-state  
energy with respect to the real variational parameters
$\lambda_{\Gamma}$.  
This minimization problem is numerically tractable because the 
number of variational parameters is now reduced to  
the dimension of the atomic Hilbert space, $D_{\rm at}={\cal O}(10^3)$
for $d$~electrons. 

The values for the electron-transfer 
parameters~$t_{\veci,\vecj}^{\sigma \sigma'}$ 
and the correlation parameters $U_{I_1,I_2}$ 
(or, equivalently, 
for $\epsilon_{\sigma \sigma'}^0(\veck) $ and 
$E_{\overline{\Gamma}}$) depend on the specific material under 
investigation. We shall specify them for nickel in Section~\ref{results}.

\subsection{Effective single-particle Hamiltonian for nickel}
\label{subsec:singleparticle}

For our studies on nickel we simplify the calculations by working with 
the correlator 
\begin{equation}
\hat{P}_{\rm G|\vecm} = \sum_{\Gamma} \lambda_{\Gamma}\hat{m}_{\Gamma|\vecm} 
\end{equation}  
in which we choose the states $|\Gamma \rangle_{\vecm}$ to be identical to
the eigenstates of the atomic Hamiltonian in~(\ref{Hlocalnew}),
$|\Gamma\rangle_{\vecm}\equiv |\overline{\Gamma}\rangle_{\vecm}$.
Therefore, our main numerical 
problem is the minimization of the variational energy with respect to 
the parameters $\lambda_{\Gamma}$. 

In order to determine the variational ground state of $\hat{H}$ one first 
needs to calculate the expectation value 
\begin{equation}  
\langle \hat{H}\rangle_{\rm G}=
E^{\rm var}(\{\lambda_{\Gamma}\},\{|\Psi_0\rangle\})    
\end{equation}  
of the Hamiltonian $\hat{H}$ as a function of the variational parameters 
$\lambda_{\Gamma}$ and the one-particle wave-function 
$|\Psi_0\rangle$. 
In the limit of infinite spatial dimensions the expectation value of the 
local Hamiltonian (\ref{Hlocalnew}) becomes
\begin{equation}  
E^{\rm at}(\{ m_{\Gamma}\})=L\sum_{\Gamma}E_{\Gamma}m_{\Gamma} \; .
\label{getmeEat}
\end{equation}  
The probability for finding the atomic state~$|\Gamma\rangle_{\vecm}$,
\begin{equation}  
m_{\Gamma}\equiv\langle \hat{m}_{\Gamma|\vecm} \rangle_{\rm G}
= \lambda_{\Gamma}^2\langle \hat{m}_{\Gamma|\vecm} \rangle_0 \; ,
\end{equation}  
may be used to replace the original variational parameters 
$\lambda_{\Gamma}$ because $\langle \hat{m}_{\Gamma|\vecm} \rangle_0$
is a simple function
of the local density-matrix for the non-interacting system,
\begin{equation}  \label{cmatrix}
C_{\sigma \sigma'}= \langle \hat{c}_{\vecm,\sigma}^+\hat{c}_{\vecm,\sigma'}
 \rangle_0\;.
\label{Cmatrix}
\end{equation}  

The one-particle product state $| \Psi_0  \rangle$ is the ground state of the 
effective single-particle Hamiltonian
\begin{eqnarray} 
\hat{T}^{\rm eff}&=&\sum_{\veck;\sigma \sigma'}
\epsilon_{\sigma \sigma'}(\veck) 
\hat{c}_{\veck,\sigma}^+\hat{c}_{\veck,\sigma'}\;,\\
\epsilon_{\sigma \sigma'}(\veck)&\equiv&\sum_{\gamma \gamma'}
Q^{\sigma \sigma'}_{\gamma \gamma'}\epsilon^0_{\gamma \gamma'}(\veck)
+\eta_{\sigma \sigma'}\;,
\end{eqnarray}  
where the tensor elements $Q^{\sigma \sigma'}_{\gamma \gamma'}$ 
are known functions of the 
parameters $m_{\Gamma}$ and the density matrix $C_{\sigma \sigma'}$, 
and $\eta_{\sigma\sigma'}$
are Lagrange parameters which act as effective on-site energies,
see Sections~\ref{gsenergy} and~\ref{qp}. The one-particle eigenvalues 
($\equiv E_{\veck,\tau}$) of $\hat{T}^{\rm eff}$ can be interpreted 
as quasi-particle energies within a Landau--Gutzwiller 
Fermi-liquid theory, see~\cite{thulpaper} and Section~\ref{qp}.
As we will see in the case of nickel, these eigenvalues
are in very good agreement with the band energies as found in ARPES
experiments. 
For the kinetic energy we find
\begin{equation} 
E^{\rm kin}(\{m_{\Gamma}\},\{C_{\sigma \sigma'}\},\{\eta_{\sigma \sigma'}\})
=\sum_{\sigma \sigma',\gamma \gamma'}
Q^{\sigma \sigma'}_{\gamma \gamma'}
\left(\{m_{\Gamma}\},\{C_{\sigma \sigma'}\}\right)
\sum_{\veck} \epsilon_{\gamma\gamma'}^0(\veck)
\langle \Psi_0 |\hat{c}_{\veck,\sigma}^+\hat{c}_{\veck,\sigma'}
| \Psi_0  \rangle \; ,
\label{ekin5}
\end{equation}  
The local density matrix $C_{\sigma \sigma'}$, via
eq.~(\ref{cmatrix}), 
and, therefore, the variational energy~$E^{\rm var}$ may be considered 
as functions of $m_{\Gamma}$ and $\eta_{\sigma \sigma'}$. The
remaining numerical task is then to 
minimize $E^{\rm var}$ with respect to these parameters.
There are, however,  still up to 
$(2N_{\rm o})^2+1$ constraints which have to be respect during the
minimization, see~(\ref{neb1})--(\ref{neb4}). 
For example, the completeness of the states $|\Gamma\rangle_{\vecm}$ 
apparently leads to the constraint $\sum_{\Gamma}m_{\Gamma}=1$. 
There is no simple recipe for the most efficient way to implement
these constraints in the numerical minimization procedure. In
previous works~\cite{PRB,EPL} we have proposed the following strategy: 
We perform a transformation of the local orbital
basis onto a new one in which the local density matrix is diagonal. 
In this new basis, the main $(2N_{\rm o})^2$ constraints could then be 
implemented by the diagonalization of a $2N_{\rm o}$-dimensional 
matrix which we called the `$Z$-matrix'. 
In the following section we present results using this strategy.
The final numerical results should not be affected by the way
the constraints are implemented.

\section{Results for ferromagnetic nickel} 
\label{results} 
 
In this section we first specify our model parameters. Then we give
some details of our variational calculations. Finally, we discuss
our results for the quasi-particle band structure and the magnetic
anisotropy.

\subsection{Model specifications} 
\label{subsec:modelspec}

The multi-band Gutzwiller theory is not an ab-initio theory. It 
is based on a multi-band Hubbard Hamiltonian whose parameters
need to be specified. For metallic systems, the use of a Hubbard
model is well justified: screening is very efficient so that the 
effects of the Coulomb interaction between electrons at distances
larger than the inverse Fermi wave number $k_{\rm F}^{-1}$ 
can be incorporated in the `bare' band structure, 
and only the local, effective Coulomb
matrix elements need to be taken into account explicitly.

\subsubsection{Electron-transfer amplitudes and local potential terms}

For nickel, we consider a minimal model which includes only 
those bands which are partly filled within a paramagnetic LDA
calculation.
Therefore, our multi-band Hubbard Hamiltonian~$\hat{H}$ 
comprises of 18~spin orbitals, namely $3d$, $4s$, and $4p$.
The non-magnetic local-density approximation to DFT provides the LDA band 
structure~$\epsilon_{\tau}^{\rm LDA}(\veck)$.
We represent the bare energy-band matrix 
elements~$\epsilon_{\sigma \sigma'}^0(\veck)$ in the
kinetic energy operator~$\hat{T}$ 
in terms of real electron-transfer integrals~$t^{\sigma \sigma'}(\vecr)$ 
in the two-center approximation~\cite{SlaterKoster} 
which range up to third nearest
neighbors, $N_{\vecr}\leq 3$. 
We choose these parameters
and the local potential terms $\epsilon_{\sigma\sigma'}$ in such a way 
that our tight-binding fit reproduces the LDA band structure. 
In this fit we include
information on the symmetry of the single-particle states~\cite{Mattheiss}.

\begin{table}[htb]
\begin{center}
\begin{tabular}[t]{r|rrrrr}
$N_{\vecr}$ & $ss\sigma$    & $sp\sigma$   &$ sd\sigma$ &$pp\sigma$&
$pp\pi$ \\
\hline
1  &  $ -1.0292$ & $ 1.2047  $ & $ -0.5933  $ &
 $ 1.2144 $ & $ -0.5284 $\\
2  &  $ -0.1039 $ & $  0.2234$ & $  -0.1089 $ & 
$  0.5989$ & $  -0.2205 $\\
3  &  $ -0.0050$ & $  -0.0223$ & $  -0.0223 $ & 
$  0.0137 $ & $  0.0076 $\\[12pt]
    & $pd\sigma$& $pd\pi$&$dd\sigma$&  $dd\pi$  &  
 $dd\delta$\\
\hline
1  & $  -0.6960 $ & $  0.2300$ & $  -0.4780 $ &
 $  0.3150$ & $  -0.0481 $\\
2  &  $ -0.2092 $ & $  0.0524$ & $  -0.0848$ & 
$   0.0336$ & $  -0.0007 $\\
3  &  $ -0.0439$ & $  -0.0023$ & $  -0.0245$ &
 $  -0.0011$ & $   0.0024 $\\
\end{tabular}
\end{center}
\caption{Electron-transfer parameters~$t_{\sigma\sigma'}(\vecr)$ to
$N_{\vecr}$-th neighbors (all energies in eV).\label{table1}}
\end{table}

For the optimal choice, the $\veck$-averaged
root-mean-square deviation between 
our tight-binding fit and the LDA band structure
up to $2\, {\rm eV}$ above the Fermi energy 
is $15 \,{\rm meV}$ for the $3d$~bands
and for the $4sp$ bands. 
Our values of the electron-transfer parameters are summarized in 
Table~\ref{table1}.
These model parameters have already been used for the calculations
reported in Refs.~\cite{EPL,Woelfle}.

For the on-site parameters we find for both spin species
\begin{equation}
\begin{array}{lllll}
\epsilon_{4s,4s} &\equiv&\epsilon_s &=& \hphantom{-}5.6022 \,{\rm eV} \; , 
\\[3pt]
\epsilon_{4p,4p}^{\rm LDA} &\equiv&
\epsilon_p^{\rm LDA} &=& \hphantom{-}8.5335 \,{\rm eV} \; ,
\\[3pt]
\epsilon_{4p,4p} &\equiv&
 \epsilon_p^{\rm shift} &=& \hphantom{-}7.7835\,{\rm eV}
\;,\\
\epsilon_{3d(t_{2g}),3d(t_{2g})}&\equiv &
\epsilon_{t_{2g}} &=&-0.0290\, {\rm eV}\; ,
\\[3pt] 
\epsilon_{3d(e_{g}),3d(e_{g})}&\equiv &
\epsilon_{e_g} 
&=&\hphantom{-}0.0436\, {\rm eV}\; .
\end{array}
 \end{equation}
where $\epsilon_p^{\rm LDA}$ is the result from the fit to LDA
and $\epsilon_p^{\rm shift}$ is used in practical calculations 
for a better agreement between our Gutzwiller theory and ARPES experiments 
for nickel, see below. For the total crystal-field splitting we find 
$\epsilon_{\rm cf}=\epsilon_{t_{2g}}-\epsilon_{e_g} =
-0.0726\, {\rm eV}$.

The energetically 
highest-lying state of pure $d$~character 
is $X_5$ (purely $t_{2g}$),
$0.18\, {\rm eV}$ above the Fermi energy. The state $X_2$ (purely
$e_{g}$) lies $0.025\, {\rm eV}$ above~$E_{\rm F}$.
The width of the $3d$~bands can be estimated from 
$E(X_5) - E(X_1) = 4.45\, {\rm eV}$ ($X_1$ is predominantly
of $d(e_g)$ character), or from 
$E(X_5) - E(L_1) = 4.63\, {\rm eV}$ ($L_1$ is predominantly
$d(t_{2g})$).

Next, we address the $L_{2'}$-state. Experiment locates this state
at about $1.0\, {\rm eV}$ below~$E_{\rm F}$.
The SDFT calculations for ferromagnetic nickel as well as
our Gutzwiller theory find the $L_{2'}$-state about
$0.3\, {\rm eV}$ below the Fermi energy when we use
the LDA $4p$ orbital energy
$\epsilon_p^{\rm LDA} = 8.5335 \,{\rm eV}$.
Therefore, we shift the $4p$~orbital energy
by $0.75\, {\rm eV}$ to $\epsilon_p^{\rm shift} = 7.7835 \,{\rm eV}$, and find
the $L_{2'}$-state $0.97\, {\rm eV}$ below~$E_{\rm F}$.
In the following we present results which use 
$\epsilon_p^{\rm shift} = 7.7835 \,{\rm eV}$.
This choice enhances the $4p$ partial density by
approximately 0.1~electron and, correspondingly, enhances the
$3d$~hole charge by the same amount so that we work with
$n_d= 8.78$~\cite{4dTM}. The remaining 1.22~valence electrons 
are about evenly split between the $4s$~level and the three~$4p$ levels.

\subsubsection{Atomic interactions}

As described above, the paramagnetic LDA 
calculation provides a first single-particle contribution
to the local Hamiltonian,
\begin{equation}
\hat{H}_{\rm at}^{(1a)} = \sum_{\vecm,\sigma} \epsilon_{\sigma\sigma}
\hat{c}_{\vecm,\sigma}^+\hat{c}_{\vecm,\sigma}  
\end{equation}
with $|\sigma\rangle=|4s,4p,3d\rangle \otimes
|\uparrow,\downarrow\rangle$.
In addition to these single-particle contributions, we include the 
spin-orbit interaction for the $3d$~electrons only, $|\sigma\rangle=
|3d\rangle\otimes |\uparrow,\downarrow\rangle$.
Hereby we restrict ourselves to the
dominant, purely atomic contributions of the form
\begin{equation}
\hat{H}_{\rm at}^{(1b)} = \sum_{\vecm,\sigma\sigma'} \frac{\zeta}{2}
\langle \sigma | \widehat{l}_x \widetilde{\sigma}_x
+ \widehat{l}_y \widetilde{\sigma}_y + \widehat{l}_z \widetilde{\sigma}_z
| \sigma'\rangle
\hat{c}_{\vecm,\sigma}^+\hat{c}_{\vecm,\sigma'} \; .
\label{spinorbitHamilt}
\end{equation}
Here, $\zeta$ is the strength of the spin-orbit coupling,
$\widetilde{\sigma}_{x,y,z}$ are the three Pauli matrices,
and $\widehat{l}_{x,y,z}$ are the Cartesian components of the
vector operator for the angular momentum. 
For the spin-orbit coupling constant we choose
$\zeta = 0.080\, {\rm eV}$ as in Refs.~\cite{Abragam,BrunoIFFJ}. 

The second part of the local interaction is the two-particle
Coulomb interaction,
\begin{equation}
\hat{H}_{\rm at}^{(2)} = \sum_{\vecm} 
\sum_{\sigma_1,\sigma_2,\sigma_3,\sigma_4}
{\cal U}(\sigma_1,\sigma_2;\sigma_3,\sigma_4)
\hat{c}_{\vecm,\sigma_1}^{+}\hat{c}_{\vecm,\sigma_2}^{+} 
\hat{c}_{\vecm,\sigma_3}\hat{c}_{\vecm,\sigma_4} \; .
\label{twoparticle}
\end{equation}
The intra-atomic interactions in the $4s$ and $4p$ shell are rather weak when
compared to the broad $4sp$~energy bands. Thus, we expect 
only small correlation effects in these bands. We also neglect
correlations between $4sp$~electrons and $3d$~electrons beyond those
contained in the LDA band structure.
This is a more serious approximation 
as we neglect magnetic polarization effects on the 
$4sp$ bands and thus may underestimate the $4sp$~contribution 
to the magnetic moment. 
Under these assumptions, the spin-orbit sum in~(\ref{twoparticle})
runs over the $3d$~orbitals only. The Hamiltonian
of the full atomic problem, 
$\hat{H}_{\rm at} = \hat{H}_{\rm at}^{(1a)}+\hat{H}_{\rm at}^{(1b)}
+\hat{H}_{\rm at}^{(2)}$, provides 
the matrix elements between two atomic configurations
$I_1$ and $I_2$ in~(\ref{Hlocalnew}),
\begin{equation}
U_{I_1,I_2}  = 
{}_{\vecm}\langle I_1 |
\hat{H}_{\rm at}^{(1a)}+\hat{H}_{\rm at}^{(1b)}+
\hat{H}_{\rm at}^{(2)} |I_2 \rangle_{\vecm} \; .
\end{equation}
The atomic eigenstates $|\Gamma\rangle_{\vecm}$
can be written as product states,
\begin{equation}
|\Gamma\rangle_{\vecm} = |\Gamma_{3d}\rangle_{\vecm}
|\Gamma_{4s}\rangle_{\vecm} |\Gamma_{4p}\rangle_{\vecm} \; .
\end{equation}
In our Gutzwiller theory we correlate
the $3d$~states~$|\Gamma_{3d}\rangle_{\vecm}$ only,
i.e., the variational parameter~$\lambda_{\Gamma}$ is independent
of the $4sp$ configuration~$|\Gamma_{4s}\rangle_{\vecm} 
|\Gamma_{4p}\rangle_{\vecm}$.
This leads to a numerically tractable problem with ${\cal O}(10^3)$
variational parameters for the atomic $3d$~states.

Naturally, we must not completely ignore the Coulomb interaction
of the $4sp$ electrons. In this case
a big charge flow from the $3d$ to the $4sp$ bands 
and unphysically small occupations of the $3d$~shell would result.
One way to overcome the charge flow problem is to introduce
a chemical potential which keeps the $3d$ partial charge fixed
during the calculations, $n_d=8.78$; 
see~\cite{Woelfle} for an alternate method
which gives essentially the same results.

Gutzwiller theory produces
an optimum magnetic (spin-only) moment which is too large by about~10\%. 
It should be mentioned that
the SDFT results show a similarly large overshooting of the magnetic moment.
To best compare with experiment,
we thus carry out calculations at fixed magnetic moment for the
total energy, with either the spin-only moment fixed to the
(spin-only) experimental value of 
$\mu_{\rm spin}=0.55 \mu_{\rm B}$~\cite{EPL} or with the total
magnetic moment fixed to the full experimental moment 
of~$\mu=0.606 \mu_{\rm B}$
when the spin-orbit coupling is included~\cite{nextpaper}.

\subsubsection{Parameters for the Coulomb interaction}

It remains to determine the Coulomb interaction parameters
${\cal U}(\sigma_1,\sigma_2;\sigma_3,\sigma_4)$
in~(\ref{twoparticle}).
In the spherical-atom approximation, there are only three 
independent interaction parameters,
namely the Slater--Condon integrals $F_0$, $F_2$, and $F_4$,
from which all Coulomb interaction parameters can be determined.
We prefer the Racah definition, see~\cite{Sugano},
where the parameters $A$, $B$, and~$C$ are used which
are linear combinations of the Slater--Condon integrals.

The spherical-atom approximation is excellent in cubic systems.
In principle, the Coulomb interaction among $3d(t_{2g})$ electrons
may differ from the Coulomb interaction among $3d(e_g)$ electrons
because the radial parts of their orbital wave functions can be different.
Measurements of $d$-$d$ transitions of magnetic impurities 
with cubic site symmetry in non-magnetic oxide hosts 
show that these differences are marginal.
The $d^3$ multiplets ${}^3 H$ and ${}^3 P$, 
which are accidentally degenerate in
spherical-atom approximation, split by cubic two-particle corrections
but not by the crystal field. However, they are found to be 
degenerate within the 
experimental resolution of some ${\rm meV}$~\cite{Sugano}.
In test calculations we have used different $A$~parameters
for $d(t_{2g})$~electrons and $d(e_g)$~electrons 
but we have not found any features in our quasi-particle energies 
which would indicate a failure of the spherical-atom approximation.

The Racah parameters $B$~and~$C$ are related to the Slater integrals 
$F_2$ and~$F_4$.
They determine the splitting of the multiplets of a specific
$3d^{n}$ configuration. They can be determined experimentally 
from $d$-$d$ transition spectra for magnetic
impurity ions in non-magnetic insulating hosts.
Typical values for all kinds of transition metal ions 
are tabulated in~\cite{Sugano}.
It is found experimentally 
that the ratio $C/B$ varies smoothly between 4~and~5;
$C/B = 4$ is obtained theoretically when hydrogen $3d$ wave functions
are used. Experimentally, the values for $B$~and~$C$ for Ni$^{2+}$
and~Ni$^{3+}$ ions are known~\cite{Sugano}. When we linearly extrapolate
these values to a neutral atom we find our values
for nickel as $B = 0.09\, {\rm eV}$ and $C = 0.40\,  {\rm eV}$ so that
we employ a ratio $C/B = 4.4$. 
The values for $B$ and $C$ are close to the `bare' atomic values, 
i.e., the screening appears to be of little importance to~$B$ and~$C$.
For a fixed ratio $C/B$,
we may replace the two interaction parameters~$B$ and~$C$
by a single effective parameter $J$, as is quite often done
in the LDA+$U$ literature.
This exchange coupling $J$ is related to $B$ and $C$ by
\begin{equation}
\frac{B}{2} + \frac{C}{5} = \frac{J}{7}\;.
\end{equation}
For our nickel values for~$B$ and~$C$ we find $J = 0.88\, {\rm eV}$,
a value very similar to the 
ones used by Anisimov et al.~\cite{LDApU} and others.

The Racah parameter~$A$ (basically the $U$~parameter of the Hubbard model)
determines the separation of the various $d^n$ multiplets. 
The `bare' values, as calculated, e.g., from
atomic wave-functions are of the order of $25\, {\rm eV}$, 
as discussed already in Herring's book on magnetism~\cite{Herring} 
and confirmed recently in~\cite{Czycholl}.
There is a technique to extract $U$~parameters using
`constraint' density-functional theory and supercell geometries.
It has been found that for minimum-basis models 
the~$U$-values are smallest. For example, in the case of the cuprates,
values $U\leq 2\, {\rm eV}$ were found for a single-band Hubbard model 
whereas a three-band model including the oxygen $2p$~states employs
$U_{dd}\approx 8\, {\rm eV}$~\cite{Hybertsen}. Similar results have
been found in the case of BaBiO$_3$~\cite{Vielsack}. 
For $3d$~band models, values of~$U=4\ldots 6\, {\rm eV}$
have been reported~\cite{Anisimov}. For the $4s$-$4p$-$3d$ multi-band
case of nickel,
no such calculations are available but values~$U\approx 10\, {\rm eV}$
appear to be reasonable.
We choose~$A$ in such a way that
our Gutzwiller theory reproduces the experimental
$3d$~band width. For our 18~orbital basis we find $A = 9\, {\rm eV}$
to reproduce best the quasi-particle $3d$~band 
width~$W=3.3\, {\rm eV}$; a value of $A=12\, {\rm eV}$ leads 
to~$W=3.0\, {\rm eV}$.
If we worked with a model of only ten $3d$~spin orbitals, the corresponding 
band-width reduction is achieved for $A\approx 4\, {\rm eV}$.

\subsection{Details of the calculations}
\label{subsec:details}

The minimization problem consists of two parts. We have to optimize
the Hubbard interaction through the variational
parameters~$m_{\Gamma}$ (`internal' variation)
and the kinetic energy through the one-particle product
state~$|\Psi_0\rangle$ and its quasi-particle excitation spectrum
(`external' variation). During the variations we keep the 
partial densities fixed through a `chemical potential' 
and work for fixed magnetization
(spin-only in the absence of spin-orbit coupling).
These two minimization steps are done recursively.
The separation of `internal' and `external' minimization offers
the advantage that computer time-consuming integrations in
momentum space are reduced to a minimum.

Experimentally, the anisotropy energy is 
$E_{\rm aniso}\approx -3\, \mu{\rm eV}$ per atom.
Therefore, our total-energy calculations must reach an accuracy
of ${\cal O}(0.1\, \mu{\rm eV})$ per atom.

\subsubsection{Band calculations}

The calculation of the kinetic energy contribution to the
variational ground-state energy requires the sum over 
quasi-particle energies up to the Fermi energy~$E_{\rm F}$.
In the absence of the spin-orbit coupling we use the point-group
symmetry of the fcc lattice and calculate all quantities in
the irreducible part of the Brillouin zone. This amounts to a reduction of
the zone volume by a factor of~48. In order to generate
our mesh in momentum space, we divide the distance between
the $\Gamma$ and the $X$~point by~20.
This is the standard $\veck$-mesh used
in all our calculations. For the calculations
with cubic symmetry this results in 916 points and
$4.0\cdot 10^3$ tetrahedra for the mapping of the Fermi surface
in the irreducible part of the Brillouin zone.
For the calculations including spin-orbit coupling, we use the same mesh.
Now that the $\veck$-summation runs over the full Brillouin zone,
we sample over $3.4\cdot 10^4$ $\veck$-points and $1.6\cdot 10^5$ tetrahedra.

For the numerical integrations required for the calculation of the
kinetic energy, we use the tetrahedron method as described
in~\cite{Rath}. Additionally, in all calculations we employ 
a refinement of the numerical integration near~$E_{\rm F}$
whereby we check whether or not a tetrahedron contains states 
within a certain energy shell around $E_{\rm F}$. If so, we divide
the tetrahedron into eight smaller ones and repeat the integration
using the smaller tetrahedra.
For the spin-only (full moment) nickel calculations our
integration refinement near~$E_{\rm F}$ results in 
$4.0\cdot 10^4$ ($1.7\cdot 10^6$) additional tetrahedra in
momentum space close to~$E_{\rm F}$.

In order to achieve an accuracy of~$10^{-7}\, {\rm eV}$ for the
ground-state energy, we need to suppress numerical noise.
For example, a numerical integration yields the elements 
of the single-particle on-site density matrix~$\widetilde{C}$~(\ref{Cmatrix}).
For cubic symmetry, some diagonal elements should be identical
but they show some scattering. Even worse, off-diagonal
elements which should be zero by symmetry are finite, of the
order ${\cal O}(10^{-6})$, albeit we perform our calculations 
in double precision. The main reason for the numerical noise
is the limited accuracy of the tetrahedral method 
and of the integration refinement.
In principle, numerical noise can be reduced 
by choosing a finer $\veck$-mesh; finite computational resources
bar this route to higher accuracy.
Instead, we take advantage of all symmetries of the problem,
e.g., in a system with cubic symmetry the on-site
density matrix~$\widetilde{C}$ becomes diagonal.
Therefore, off-diagonal matrix elements can be set to zero, and
diagonal elements are replaced by their average.

\subsubsection{Variational procedure}

The variational energy has to be minimized with respect to the 
internal variational
parameters~$m_{\Gamma}$ and the external parameters 
$\eta_{\sigma\sigma'}$. The latter can be interpreted 
as the effective on-site energies,
crystal-field splittings, exchange splittings, and spin-orbit couplings.
As shown in Section~\ref{subsec:modelspec}
for cubic nickel, symmetry allows three external variational parameters
for the $3d$~electrons:
the effective crystal-field splitting~$\epsilon^{\rm eff}_{\rm cf}$
and the effective exchange splittings 
$\Delta^{\rm eff}(t_{\rm 2g})$ and $\Delta^{\rm eff}(e_{\rm g})$,
which are different for $3d(t_{2g})$ and $3d(e_g)$ electrons.
Symmetry also allows exchange splittings for the $4sp$~orbitals 
but these turn out to be of very minor 
importance, as long as we do not include the atomic electron-electron 
interaction between $3d$~electrons and $4sp$~electrons, 
see Section~\ref{subsec:modelspec}.
When we include the spin-orbit coupling~(\ref{spinorbitHamilt}) 
there is another external variational parameter, $\zeta^{\rm eff}$, 
the magnitude of the spin-orbit coupling for our quasi-particle
description. 

The optimization of the external parameters for a given tensor~$\widetilde{Q}$
is analogous to the Hartree--Fock variational theory,
i.e., it is a familiar self-consistent band-structure problem.
We would reproduce the results of a pure mean-field LDA+U approach if we 
set $Q_{\gamma \gamma'}^{\sigma \sigma'}=\delta_{\gamma,\sigma}
\delta_{\gamma',\sigma'}$. In the case of nickel, however, we found that 
 $\widetilde{Q}$ deviates significantly from this simple diagonal
form. This explains why mean-field theories fail to describe the 
electronic properties of nickel correctly.

The variation of the internal variational parameters $m_{\Gamma}$
was the centerpiece of our programming efforts. 
All our schemes use random number generators for the starting
values of the variational parameters so that we can easily
test the stability of our results.
We repeat the runs for different seeds and study the scatter of
the results, not only in the total energy, but also 
in the values of the individual variational parameters. 

For fixed external variational parameters, the kinetic energy~(\ref{ekin5})
changes only due to the internal variational parameters~$m_{\Gamma}$.
The tensor $\widetilde{Q}$ in~(\ref{ekin5}) is a product of 
matrices~$\widetilde{q}$ whose entries can be written in the form
\begin{equation}
q_{\sigma_1,\sigma_2}=\sum_{\Gamma_1,\Gamma_2} 
\Sigma(\sigma_1,\sigma_2;\Gamma_1,\Gamma_2)
\sqrt{m_{\Gamma_1}}\sqrt{m_{\Gamma_2}}\;.
\end{equation}
Most of the elements of the tensor~$\widetilde{\Sigma}$ 
can be prepared before we start
the variation of the internal variational parameters,
i.e., they do not change during the optimization of the 
parameters~$m_{\Gamma}$. Note that
the eigenvalues and eigenvectors of the matrix~$\widetilde{Z}$ 
feed into the tensor $\widetilde{\Sigma}$.
The matrix~$\widetilde{Z}$ 
of size $10\times 10$ has to be
recalculated and diagonalized at each internal variation step
as it ensures fulfillment of the conditions~(\ref{neb1}) and~(\ref{neb2}),
see Section~\ref{subsec:conditions}. However, only a few
of the elements in~$\widetilde{\Sigma}$ are affected by these recalculations.

The present code simply varies each $m_{\Gamma}$ one by one, 
calculates the matrix~$\widetilde{Z}$ and the 
corrections to the matrix~$\widetilde{q}$ in each step, 
computes the matrices~$\widetilde{q}$ and~$\widetilde{C}$,
and carries out the total-energy summations both for the interaction energy
and the kinetic energy.
The use of the matrix~$\widetilde{q}$ considerably speeds up
the calculation and variation of~$\widetilde{Q}$ and thus of the
kinetic energy.

A large number of variational parameters is very small, e.g.,
the probability~$m_{\Gamma}$ for a local configuration Ni$^{7+}$
is tiny. We use a cut-off below which such 
variational parameters are set to zero and are no longer varied.
In general, big variational parameters influence
the energy more strongly than small ones
so that we vary big variational parameters more often.
Typically, we perform $1.0\cdot 10^5$ to $5.0\cdot 10^5$ variations
to reach convergence. Hereby, the energy reduction 
in the last $2.0 \cdot 10^4$ variations
is less than $0.1\, \mu{\rm eV}$. When we repeat the run for different
seeds, the total energy scatters by less than
 $1\, \mu{\rm eV}$. The individual scatter of band and
interaction energies is of the order~${\cal O}(\mu{\rm eV})$. The scatter in 
the most significant $m_{\Gamma}$ values is of the order ${\cal O}(10^{-4})$, 
the same holds true for the diagonal elements of the matrix~$\widetilde{q}$.

\subsection{Quasi-particle band structure without spin-orbit coupling} 
\label{subsec:qpbands}

\begin{figure}[htb]
\centerline{\includegraphics[height=9.5cm]{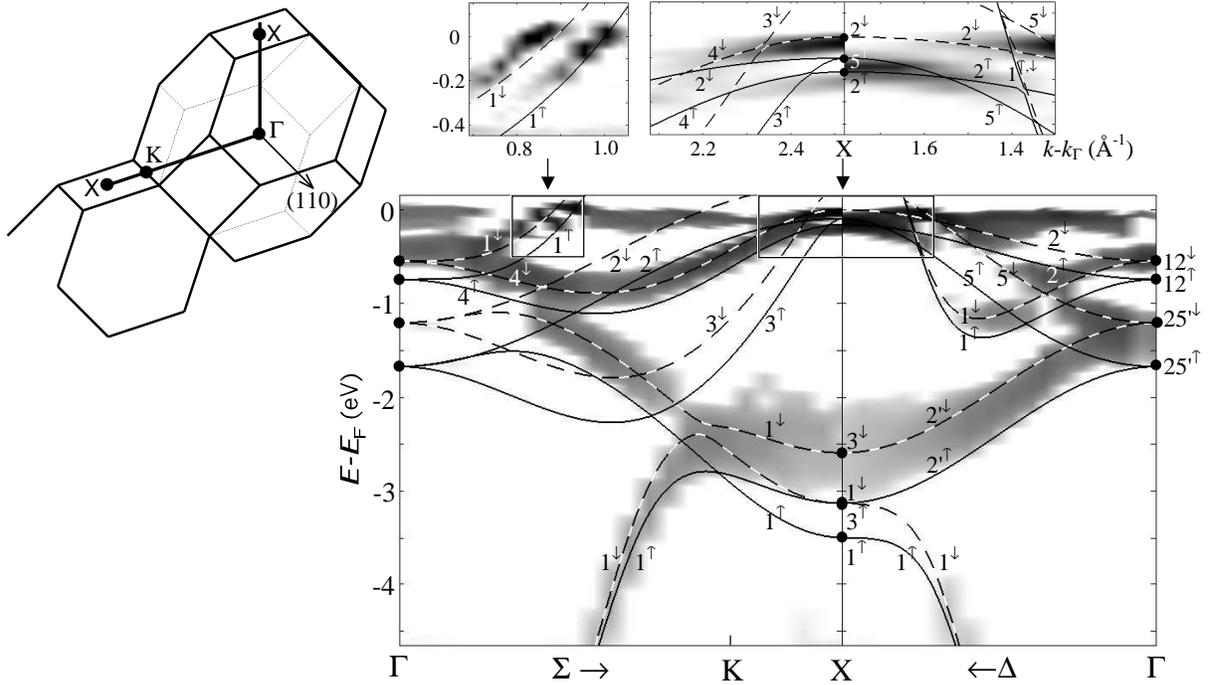}}
\caption{Grey-scale plot of the negative second derivative 
of the ARPES intensity for nickel with respect to energy, 
$-d^2I/dE^2$, on a logarithmic scale (insets:
linear scale) for the $\Gamma KX$ and $\Gamma X$
directions of the Brillouin zone. 
The dispersionless structure at $E_{\rm F}$ is due to a residual 
Fermi edge resulting from indirect transitions. Some bands
($\Delta_2$, $X_1$, $\Sigma_2$, $\Sigma_3$,
$\Sigma_1$ towards $\Gamma$) are not seen
due to unfavorable matrix elements, depending on geometry
and chosen final state~\protect\cite{EPL}. 
Theoretical curves are G-DFT, see
table~\protect\ref{table2}.\label{fig1}}
\end{figure}

First we address the case where we set the spin-orbit coupling to zero
and work with a fixed spin-only moment of $\mu_{\rm spin}=0.55$.
Figure~\ref{fig1} and table~\ref{table2} summarize 
our results which are in very good agreement with ARPES experiments.

Our calculations show the expected dependence of the $3d$~band width on the
magnitude of the Racah parameter~$A$ via the hopping 
reduction tensor $\widetilde{Q}$. 
As can be seen from table~\ref{table2}, the $3d$~band width agrees 
very well with experiment for $A = 9\, {\rm eV}$.
The Racah parameters only indirectly influence the quasi-particle bands through
the matrix~$\widetilde{Q}$ whose elements determine the reduction
of the electron-transfer amplitudes. 
Modifications of the values for $A$, $B$, and~$C$ cause overall 
energy shifts but they only weakly influence specific band states.
Now that the value of~$A$ is so much bigger than~$B$ and~$C$,
there is very little dependence of the $3d$~band width
on the value of the Racah parameters~$B$ and~$C$. 

A central result of our Gutzwiller theory is the anisotropic
exchange splitting. The magnetization of our sample
leads to the effective exchange splittings
for $3d(e_g)$~states and $3d(t_{2g})$~states,
$\Delta^{\rm eff}(t_{2g}) = 0.25\, {\rm eV}$, and 
$\Delta^{\rm eff}(e_g)= 0.11\, {\rm eV}$ which
differ by more than a factor of two.
Correspondingly, in the quasi-particle band structure 
near $E_{\rm F}$, states of pure $t_{2g}$ character (e.g., $X_5$, $W_{1'}$)
exhibit a large splitting of majority and minority spin states 
of about $400\, {\rm meV}$ whereas
the states with pure $e_g$ character (e.g., $X_2$) exhibit small splittings
of about $150\, {\rm meV}$ to $200\, {\rm meV}$, see table~\ref{table2}.
An important consequence of this anisotropy is the result
that the state $X_{2\downarrow}$ lies slightly 
below the Fermi energy. This result is robust in all our calculations
with Gutzwiller theory, e.g, it also applies for $A=12\, {\rm eV}$.

The large anisotropy of the exchange splitting has been found
experimentally by spin-polarized ARPES studies along the $\Lambda$-line
of the Brillouin zone~\cite{Kaem90}, and is well reproduced by our 
Gutzwiller theory, see table~\ref{table2}. In contrast, the SDFT energy 
bands show a much larger but rather isotropic exchange splitting because
the exchange-correlation potential is constructed both
from superposed $3d(t_{2g})$ and $3d(e_g)$ charge densities, i.e.,
the exchange-correlation potential is independent of the orbitals. 
In our Gutzwiller theory
the two values for the exchange splittings result 
from two independent variational parameters.

It is relatively easy to understand the reason 
for the large anisotropy of the exchange splittings. 
The width of the $3d(t_{2g})$ bands is mainly determined 
by the first-neighbor electron-transfer integral $(dd\sigma)_1$, the largest
$3d$-$3d$ matrix element, see table~\ref{table1}.
The width of the $3d(e_g)$~bands does not depend on the 
$(dd\sigma)_1$~integral,
but only on the much smaller $(dd\sigma)_2$ integral to second
nearest neighbors. In addition there is some first-neighbor 
$3d(e_g)$-$3d(t_{2g})$ hybridization via the large 
$(dd\pi)_1$ integral.
Thus, the most strongly anti-bonding bands are the $3d(t_{2g})$ bands 
which form the top of the $3d$~bands. 
Electron-electron interaction enforces the
exchange splitting of the bands up to the point where the limit of strong
ferromagnetism is reached and the majority spin bands are
completely filled. In the minority spin bands, some energy can be
gained when the occupation of the anti-bonding $3d(t_{2g})$ bands 
is reduced, at the expense of the occupation of the top of the 
$3d(e_g)$ bands, i.e., of the occupation of the $X_{2\downarrow}$ state. 

\vfill
\begin{table}[hb]
\begin{center}
{\small \begin{tabular}{lclcll}
 Symmetry & Character & Experiment & Reference
 & G-DFT &  SDFT\\
\hline
\mbox{}&&&&&\\[-6pt]
$\langle \Gamma_1 \rangle$   & S &  8.90$\pm$0.30 & \protect\cite{EPL} 
& 8.86 & \hphantom{$-$}8.96[$-$0.11]\\
$\langle \Gamma_{25'}\rangle$& T & 1.30$\pm$0.06 & \protect\cite{EPL}
& 1.44[0.46] & \hphantom{$-$}1.99[0.43]\\
$\langle \Gamma_{12}\rangle$ & E & 0.48$\pm$0.08 & \protect\cite{EPL}
& 0.65[0.195] & \hphantom{$-$}0.86[0.41]\\
$\langle X_1 \rangle$        & sE& 3.30$\pm$0.20 & \protect\cite{EP80}
& 3.31[0.36] & \hphantom{$-$}4.37[0.20]\\
$\langle X_3 \rangle$        & T & 2.63$\pm$0.10 & \protect\cite{EPL}
& 2.86[0.54] &\hphantom{$-$}3.82[0.37]\\
$ X_{2\uparrow}$             & E & 0.21$\pm$0.03 & \protect\cite{EPL}
& 0.165 & \hphantom{$-$}0.35\\
$ X_{2\downarrow}$           & E & 0.04$\pm$0.03 & \protect\cite{EPL}
& 0.01 & $-$0.09\\
$ X_{5\uparrow}$             & T & 0.15$\pm$0.03 & \protect\cite{EPL}
&  0.10 & \hphantom{$-$}0.23\\
$\Delta_{e_g}(X_2)$          & E & 0.17$\pm$0.05 & \protect\cite{EPL}
&  0.155& \hphantom{$-$}0.44\\
$\Delta_{t_{2g}}(X_5)$       & T & 0.33$\pm$0.04 & \protect\cite{EHK78}
& 0.38 & \hphantom{$-$}0.56\\
$\langle K_1 \rangle$       &sptE& 3.10$\pm$0.20 & \protect\cite{EP80}
& 2.76[0.33] &\hphantom{$-$}3.66[0.26]\\
$\langle K_2 \rangle$       &spTe& 2.48$\pm$0.06 & \protect\cite{EP80}
&  2.59[0.50] &\hphantom{$-$}3.37[0.32]\\
$\langle K_3 \rangle$        & pT& 0.90$\pm$0.20 & \protect\cite{EP80}
& 1.36[0.41] &\hphantom{$-$}1.73[0.37]\\
$\langle K_4 \rangle$        & pE& 0.47$\pm$0.03 & \protect\cite{EPL}
& 0.51[0.185] & \hphantom{$-$}0.70[0.41]\\
%
%
$\langle L_1 \rangle$        & sT& 3.66$\pm$0.10 & \protect\cite{EPL}
&3.51[0.515] & \hphantom{$-$}4.56[0.23]\\
$\langle L_3 \rangle$        & tE& 1.43$\pm$0.07 & \protect\cite{EPL}
&1.51[0.34] &  \hphantom{$-$}2.02[0.40]\\
$ L_{3\uparrow}$             &Te & 0.18$\pm$0.03 & \protect\cite{EPL}
&0.22 &  \hphantom{$-$}0.38[0.50]\\
$\langle L_{2'}\rangle$      & P & 1.00$\pm$0.20 & \protect\cite{EP80}
& 0.97[0.0] & \hphantom{$-$}0.24[$-$0.12]\\
$\langle W_{2'}\rangle$      & pE& 2.60$\pm$0.20 & \protect\cite{EP80}
& 2.66[0.31] & \hphantom{$-$}3.46[0.24]\\
$\langle W_3 \rangle$        & pT& 1.70$\pm$0.20 & \protect\cite{EP80}
& 2.04[0.47]& \hphantom{$-$}2.69[0.36]\\
$\langle W_1 \rangle$        & sE& 0.65$\pm$0.10 & \protect\cite{EP80}
& 0.69[0.20] & \hphantom{$-$}0.94[0.39]\\
$ W_{1'\uparrow}$            & T & 0.15$\pm$0.10 & \protect\cite{EP80}
& 0.11 & \hphantom{$-$}0.23[0.56]\\
$\langle\Lambda_{3;1/3}\rangle$&ptE&0.57[0.16$\pm$0.02]& \protect\cite{Kaem90}
& 0.67[0.22] & \hphantom{$-$}0.90[0.42]\\
$\langle\Lambda_{3;1/2}\rangle$&ptE&0.50[0.21$\pm$0.02]&\protect\cite{Kaem90}
& 0.55[0.26] & \hphantom{$-$}0.76[0.44]\\
$\langle\Lambda_{3;2/3}\rangle$&pTE&0.35[0.25$\pm$0.02]&\protect\cite{Kaem90}
& 0.33[0.29] & \hphantom{$-$}0.49[0.48]
\end{tabular}}
\end{center}
\caption{\small Binding energies in~${\rm eV}$ 
with respect to the Fermi energy~$E_{\rm F}$
($>0$ for occupied states). $\langle \ldots \rangle$ indicates the spin
average, error bars in the experiments without spin resolution
are given as $\pm$. 
Theoretical data show the spin average and the exchange splittings in square
brackets. The second column denote the orbital character of the states,
$t\equiv t_{2g}$, $e\equiv e_g$,
capital letters: dominant character. The spin-polarized data 
$\langle \Lambda_{3;f}\rangle$
were taken at fractions~$f$ of the $\Gamma$$L$ distance,
with the emphasis on the analysis of the exchange splittings.
G-DFT and SDFT calculations are without spin-orbit coupling and a fixed
spin-only moment of $\mu_{\rm spin}=0.55 \mu_{\rm B}$.
\label{table2}}
\end{table}

\newpage

\subsection{Magnetic anisotropy} 
\label{subsec:magneticanisotropy}

We now turn on the spin-orbit coupling.
Again, we perform fixed-moment calculations, now with 
the full magnetic moment $\mu = 0.606\, \mu_{\rm B}$.
Our total-energy calculations including spin-orbit coupling
are two to three orders of magnitude more time consuming than
the calculations discussed in Section~\ref{subsec:qpbands} because 
our $\veck$-sums run over the full Brillouin zone,
all hundred elements of the matrix~$\widetilde{q}$ must be calculated
and all thousand internal variational parameters must be
optimized, and additional external variational parameters
appear, e.g., the effective spin-orbit coupling~$\zeta^{\rm eff}$.

\begin{table}[htb]
\begin{center}
\begin{tabular}{c|c}
\begin{tabular}[t]{lcll}
\mbox{}&&&\\
 Symm. & Char. 
 & G-DFT &  G-DFT(SO)\\
\hline
\mbox{}&&&\\[-6pt]
$ \Gamma_{25'\uparrow}$ & T
& \hphantom{$-$}1.67 & 
\begin{tabular}{l}
\hphantom{$-$}1.74\\
\hphantom{$-$}1.67\\
\hphantom{$-$}1.17
\end{tabular}
\\
\mbox{}&&&\\[-9pt]
$ \Gamma_{25'\downarrow}$&T
& \hphantom{$-$}1.21 & \begin{tabular}{l}
\hphantom{$-$}1.25\\
\hphantom{$-$}1.21\\
\hphantom{$-$}1.17
\end{tabular}
\\
\mbox{}&&&\\[-9pt]
$X_{2\uparrow}$&E
& \hphantom{$-$}0.17 &\begin{tabular}{l}
\hphantom{$-$}0.19
\end{tabular}
\\
\mbox{}&&&\\[-9pt]
$X_{2\downarrow}$&E
& \hphantom{$-$}0.01 &\begin{tabular}{l}\hphantom{$-$}0.02 \end{tabular}
\\
\mbox{}&&&\\[-9pt]
$X_{5\uparrow}$&T
& \hphantom{$-$}0.10 &\begin{tabular}{l}
\hphantom{$-$}0.13\\
\hphantom{$-$}0.11
\end{tabular}
\\
\mbox{}&&&\\[-9pt]
$X_{5\downarrow}$&T
& $-$0.28 &\begin{tabular}{l}
$-$0.27 \\
$-$0.32
\end{tabular}
\\
\mbox{}&&&\\[-9pt]
$L_{3\uparrow}$&Te
& \hphantom{$-$}0.22 &\begin{tabular}{l}
\hphantom{$-$}0.26 ; 0.24 \\
\hphantom{$-$}0.20 ; 0.22
\end{tabular}
\end{tabular}
&
\begin{tabular}[t]{lcll}
\mbox{}&&&\\
 Symm. & Char. 
 & G-DFT &G-DFT(SO)
\\
\hline
\mbox{}&&&\\[-6pt]
$\Lambda_{3;1/3\uparrow}$&ptE
& 0.78 &\begin{tabular}{l}
0.81 ; 0.80\\
0.76 ; 0.78
\end{tabular}
\\
\mbox{}&&&\\[-9pt]
$\Lambda_{3;1/3\downarrow}$&ptE
& 0.57 &\begin{tabular}{l}
0.61 ; 0.59\\
0.56 ; 0.57
\end{tabular}
\\
\mbox{}&&&\\[-9pt]
$\Lambda_{3;1/2\uparrow}$&ptE
& 0.68 &\begin{tabular}{l}
0.72 ; 0.70\\
0.66 ; 0.69
\end{tabular}
\\
\mbox{}&&&\\[-9pt]
$\Lambda_{3;1/2\downarrow}$&ptE
& 0.42 &\begin{tabular}{l}
0.47 ; 0.44\\
0.39 ; 0.41
\end{tabular}
\\
\mbox{}&&&\\[-9pt]
$\Lambda_{3;2/3\uparrow}$&pTE
& 0.47 &\begin{tabular}{l}
0.52 ; 0.50\\
0.45 ; 0.48
\end{tabular}
\\
\mbox{}&&&\\[-9pt]
$\Lambda_{3;2/3\downarrow}$&pTE
& 0.19 &\begin{tabular}{l}
0.22 ; 0.20\\
0.16 ; 0.18
\end{tabular}
\end{tabular}
\end{tabular}
\caption{Theoretical binding energies in~${\rm eV}$ 
with respect to the Fermi energy~$E_{\rm F}$
($>0$ for occupied states) with and without spin-orbit coupling.
The second column denote the orbital character of the states,
$t\equiv t_{2g}$, $e\equiv e_g$,
capital letters: dominant character. 
For the calculations without spin-orbit coupling, G-DFT, 
we used a spin-only moment of $\mu_{\rm spin}=0.55\, \mu_{\rm B}$. In the 
spin-orbit calculation, G-DFT(SO),  
we worked with a fixed moment of $0.606\,  \mu_{\rm B}$
along the (111)-direction. The directions~$\Lambda$ and the
$L$-points in the Brillouin zone are not equivalent: the first and 
the second energy belong to the directions parallel and perpendicular to 
(111), respectively.
\label{table3}}
\end{center}
\end{table}

Our optimal ground state displays an
orbital moment of $\mu_{\rm orb}=0.052\, \mu_{\rm B}$ which agrees
very well with the experimental value of 
$\mu_{\rm orb}=0.0507\, \mu_{\rm B}$~\cite{Wohlfarth}. 
In the Hartree--Fock limit of our method, $\lambda_{\Gamma}\equiv 1$,
we obtain $\mu_{\rm orb, \, HF}=0.032\, \mu_{\rm B}$. 
Obviously, the orbital-moment contribution to the total moment almost
doubles when we apply our Gutzwiller theory. This is due to the fact that
correlations induce non-diagonal elements
in the matrix~$\widetilde{q}$; they become nonzero because the spin-orbit
coupling lifts the cubic symmetry. 
An analysis of the quasi-particle bands shows
that the non-diagonal elements of the matrix~$\widetilde{q}$
induce electron transfers with spin flips 
in the effective single-particle Hamiltonian.
These processes give an additional contribution to the orbital moment. 
The effective value for the spin-orbit coupling
becomes smaller, $\zeta^{\rm eff} = 0.070\, {\rm eV}$ for the optimum 
ground-state energy. This is a $12\%$ reduction compared to the bare
spin-orbit coupling constant $\zeta=0.080\, {\rm eV}$.
Calculations with a fixed $\zeta^{\rm eff} = \zeta$
yield higher ground-state energies and an approximately $10\%$ larger
orbital moment.

The central external variational parameters, the effective exchange
splittings $\Delta^{\rm eff}(t_{2g})$ and $\Delta^{\rm eff}(e_g)$
as well as the effective crystal-field $\epsilon^{\rm eff}_{\rm cf}$,
are very similar to the spin-only calculations. 
Consequently, the quasi-particle bands do not
change very much from the quasi-particle bands as described
in Section~\ref{subsec:qpbands}.
However, degeneracies at high-symmetry points and lines are lifted.
In table~\ref{table3}, we give some of the energies 
at high-symmetry points and lines, taken
from a calculation where the magnetic moment lies parallel to the 
$(111)$-direction, the easy axis of nickel. Near the Fermi energy 
the splittings should be detectable by high resolution ARPES. 
Note that the state which 
derives from the $X_{2\downarrow}$-state still lies slightly 
below the Fermi energy. 

\begin{figure}[bh]
\begin{center}
{\tabcolsep=2pt
\begin{tabular}[t]{@{}lcr@{}}
\includegraphics[height=7.5cm]{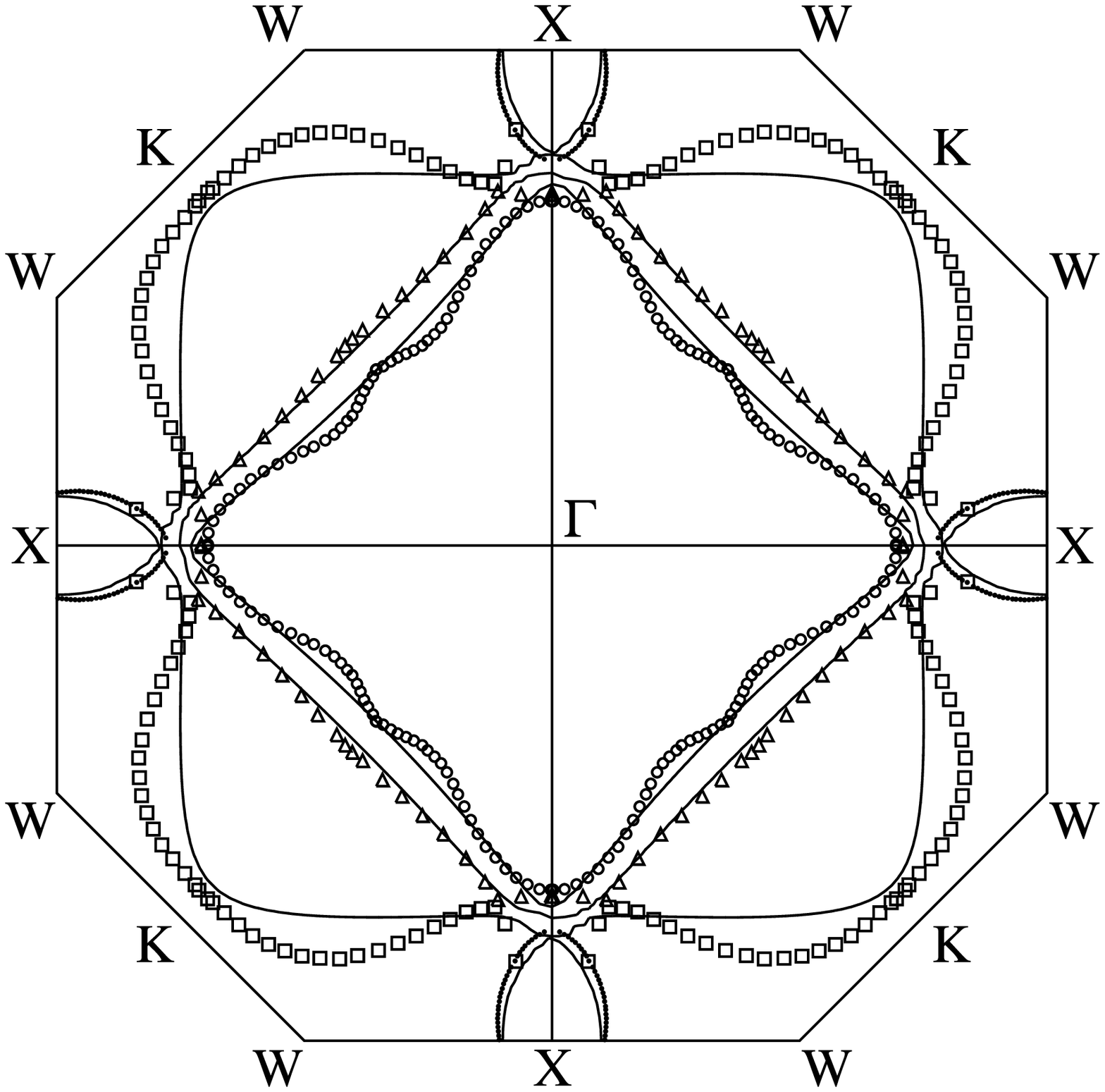}
&\mbox{} & \includegraphics[height=7.5cm]{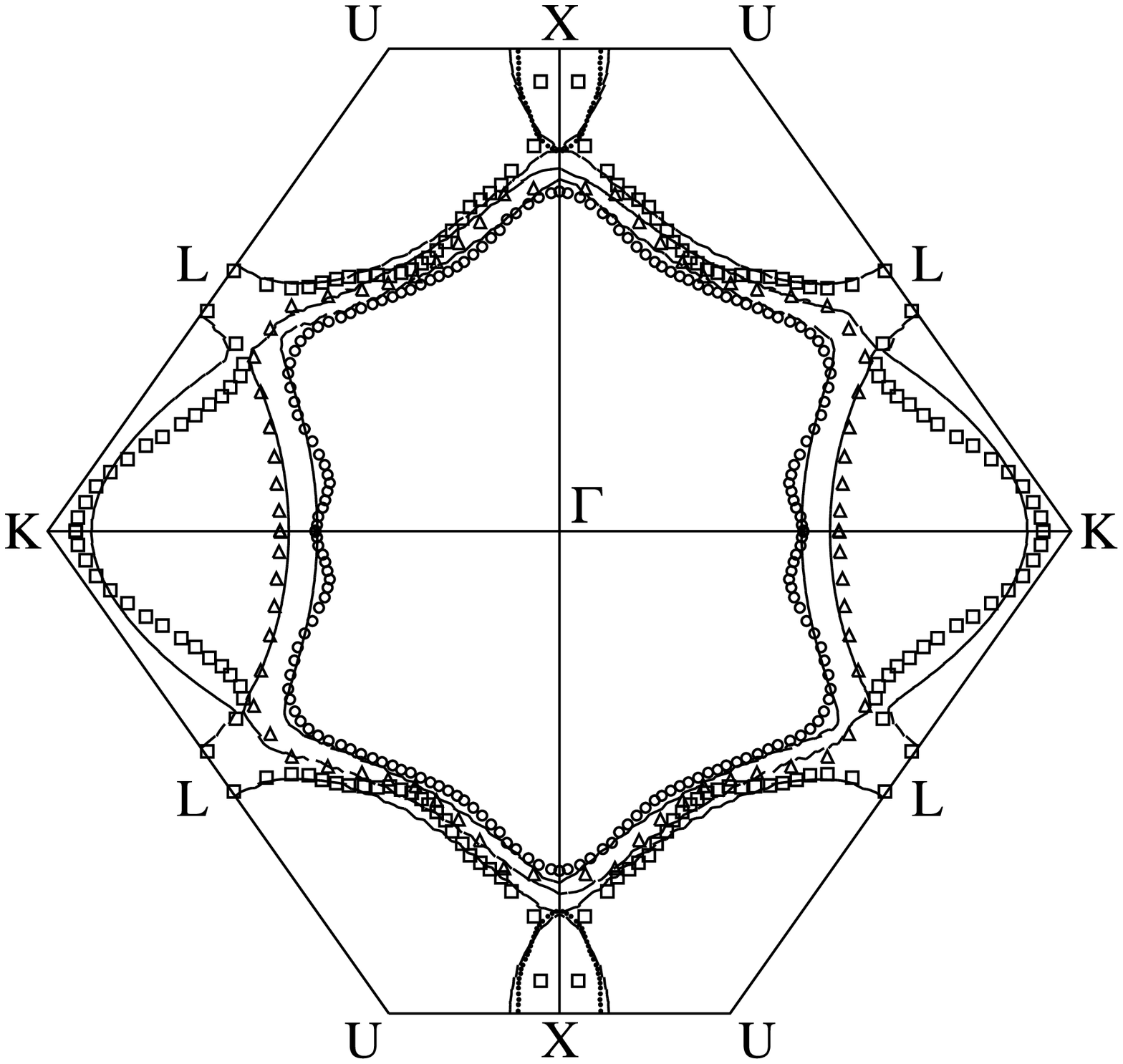}
\end{tabular}}
\caption{Cut of the Fermi surface with the plane $\Gamma XK$
and the plane $\Gamma XL$ in the Brillouin zone.
Open symbols are data of Stark, communicated in~\protect\cite{Callaway},
filled dots are de-Haas--van-Alphen data from~\protect\cite{Tsui},
and lines are the results from Gutzwiller theory.
\label{fig2}}
\end{center}
\end{figure}

The spin-orbit coupling causes the former majority and minority spin bands
to couple so that band crossings of spin-up bands and spin-down bands
are avoided. As a consequence, the topology of the Fermi surface changes
slightly. These changes affect only a few areas on the Fermi surface.
In figure~\ref{fig2},
we show a cut of the Fermi surface in the $\Gamma KX$ plane
and in the $\Gamma XL$ plane, respectively.

The theoretical Fermi surface agrees very well with the 
de-Haas--van-Alphen data of Stark,
which have been communicated through the paper of Callaway 
and Wang~\cite{Callaway}. 
However, the experimental data 
exhibit oscillations in these cuts which are not reproduced by our 
Gutzwiller theory. In fact, no band calculation whatsoever 
has produced these wiggles. We suspect that
these wiggles are artifacts of the reconstruction procedure 
from the raw data: in a magnetic field of several Tesla,
the Fermi surface topology itself can change, as we now discuss.

When we align the magnetic moment along the $(001)$-direction, the 
quasi-particle energy of the $X$-states at (100) and at~(010) 
differ from the energy of the (001)~$X$-state;
the three states are degenerate when the moment is along the (111)~direction.
The spin-orbit coupling induces a splitting of the formerly degenerate
$X_{2\downarrow}$~state by $30\, {\rm meV}$,
and the energy of the (001)~$X_{2\downarrow}$-state in the presence
of the spin-orbit coupling lies approximately
$6\, {\rm meV}$ {\sl above\/} the Fermi energy
whereas the other two state lie
$25\, {\rm meV}$ below~$E_{\rm F}$.
Thereby our Gutzwiller theory confirms 
the scenario put forward by Gersdorf~\cite{Gersdorf}.

The differences in ground-energies of the states with the magnetic moment
aligned in the (111)~direction and
with the magnetic moment aligned in the (001)~direction, 
$E_{\rm aniso}\equiv E_{111}-E_{001}$,
is $E_{\rm aniso, \, Gutz}\approx -10\, \mu{\rm eV}$ per atom.
Thus we get the correct sign and order of magnitude
of the experimental energy difference,
$E_{\rm aniso}=-3\, \mu{\rm eV}$ per atom.
In order to improve the confidence in our results,
we have repeated our numerical integrations over the Brillouin zone
with a denser mesh but for fixed external variational parameters.
Our calculations using
a denser mesh with up to $2.5\cdot 10^5$ $\veck$-points 
in the Brillouin zone confirm the result of a
$(111)$~easy axis in nickel. Therefore,
we are confident that Gutzwiller theory calculations produce
reliable results for the magnetic anisotropy energies.
Further studies with even larger $\veck$-meshes are 
in progress.

\section{Diagrammatic approach}
\label{GWFevaluate}

In this section we develop a general scheme to evaluate  
expectation values for single-site and two-site operators 
using the variational analog of Feynman diagrams.  

\subsection{Conditions} 
\label{subsec:conditions}
  
We need to evaluate expectation values of operators $\hat{O}_{\veci}$  
and $\hat{O}_{\veci,\vecj}=\hat{c}_{\veci,\sigma}^{(+)}  
\hat{c}_{\vecj,\sigma}^{(+)}$ for $\veci\neq \vecj$ with the wave  
function~(\ref{defwavefunction}),  
\begin{eqnarray}  
\langle \hat{O}_{\veci} \rangle_{\rm G} &=&  
\frac{ \langle \Psi_0 |   
\left[\prod_{\vecm \neq \veci} \hat{P}_{\rm G|\vecm}^2\right]  
\hat{P}_{\rm G|\veci}\hat{O}_{\veci}\hat{P}_{\rm G|\veci} |  
\Psi_0\rangle  
}{  
\langle \Psi_0 | \prod_{\vecm} \hat{P}_{\rm G|\vecm}^2 | \Psi_0\rangle  
} \; ,  
\label{Oi}  
\\  
\langle \hat{O}_{\veci, \vecj} \rangle_{\rm G} &=&  
\frac{ \langle \Psi_0 |   
\left[\prod_{\vecm \neq \veci,\vecj} \hat{P}_{\rm G|\vecm}^2\right]  
\hat{P}_{\rm G|\veci}\hat{P}_{\rm G|\vecj}  
\hat{O}_{\veci,\vecj}\hat{P}_{\rm G|\veci} \hat{P}_{\rm G|\vecj}  
|\Psi_0\rangle  
}{  
\langle \Psi_0 | \prod_{\vecm} \hat{P}_{\rm G|\vecm}^2 | \Psi_0\rangle  
}\; .  
\label{Oij}  
\end{eqnarray}  
The square of the local Gutzwiller correlator can be reduced to  
\begin{equation}  
 \hat{P}_{\rm G|\vecm}^2 = \sum_{I_1,I_2}   
\overline{ \lambda}_{I_1,I_2|\vecm}    
|I_1\rangle_{\vecm} {}_{\vecm}\langle I_2|  
\end{equation}  
with  
\begin{equation}  
\overline{ \lambda}_{I_1,I_2|\vecm}  
= \sum_{I'}\lambda_{I_1,I'|\vecm}\lambda_{I',I_2|\vecm} \; .  
\end{equation}  
 We aim at a diagrammatic calculation of $\langle \hat{O}_{\veci}   
\rangle_{\rm G}$ and $\langle \hat{O}_{\veci, \vecj}\rangle_{\rm G}$.  
In such an approach, the sites $\vecm\neq \veci,\vecj$   
in~(\ref{Oi}) and~(\ref{Oij}) will play the role of inner vertices.  
 
In a first step, we eliminate all local contributions at the inner  
vertices which arise after the application of Wick's theorem  
to the numerators in~(\ref{Oi}) and~(\ref{Oij}).  
As in a previous work~\cite{PRB}, we expand $ \hat{P}_{\rm G|\vecm}^2$  
in terms of new local operators,  
\begin{eqnarray}  
\hat{P}_{\rm G|\vecm}^2 &=& 1 +  \left[\hat{P}_{\rm G|\vecm}^2\right]^{\rm HF}  
\nonumber \; ,\\   
\left[\hat{P}_{\rm G|\vecm}^2\right]^{\rm HF} &=&  
\sum_{I_1,I_2;|I_1|+|I_2|\geq 2}   
x_{I_1,I_2|\vecm} \left(  
\hat{n}_{I_1,I_2,\vecm}   
- \Bigl[\hat{n}_{I_1,I_2,\vecm}      
\Bigr]^{\rm HF}  
 \right)  \; ,  
\label{PGsqnew}  
\end{eqnarray}  
where 
\begin{equation} 
\hat{n}_{I_1,I_2,\vecm}\equiv \prod_{\sigma_{1} \in I_1} 
\hat{c}_{\vecm,\sigma_{1}}^{+} 
\prod_{\sigma_{2} \in I_2}\hat{c}_{\vecm,\sigma_{2}}\;. 
\end{equation} 
The Hartree--Fock operator for an even number $n\geq 4$ of Fermion  
operators is defined recursively,  
\begin{eqnarray} \label{hfop} 
\left[  
\hat{a}_1\ldots \hat{a}_n\right]^{\rm HF} 
&\equiv&\left \langle 
\hat{a}_1\ldots \hat{a}_n 
\right\rangle_{0}\\\nonumber 
&&+\mathop{{\sum}'}_{\{\gamma_1,...,\gamma_{n}\}=0}^{1} 
(-1)^{f_{\rm s}(\{\gamma_{i}\})} 
\left\{ 
\left( 
\prod_{\ell=1}^{n}\hat{a}_{\ell}^{\gamma_{\ell}} 
\right)- 
\left[ 
\prod_{\ell=1}^{n}\hat{a}_{\ell}^{\gamma_{\ell}} 
\right]^{\rm HF} 
\right\} 
\left\langle 
\prod_{\ell=1}^{n}\hat{a}_{\ell}^{1-\gamma_{\ell}} 
\right\rangle_{0} 
\end{eqnarray}  
with  
\begin{eqnarray} 
 f_{\rm s}(\{\gamma_{i}\})&\equiv&
\sum_{\ell=1}^{n}(\ell-\frac{1}{2})\gamma_{\ell}  \; . 	 
\end{eqnarray} 
The prime in~(\ref{hfop}) indicates that  
$2\leq \sum_{\ell=1}^{n}\gamma_{\ell}\leq n-2$ is even. 
For $n=2$ we have 
\begin{equation} 
\left[ \hat{a}_1  \hat{a}_{2} \right]^{\rm HF}= 
\langle \hat{a}_1\hat{a}_2\rangle_0  \; . 
\end{equation} 
By construction, the operator in (\ref{PGsqnew}) generates  
diagrams with 
exactly $|I_1|$ ($|I_2|$) non-local lines that enter (leave) the  
lattice site~$\vecm$. All diagrams with trivial local Hartree--Fock bubbles 
are automatically excluded.  
 
For our diagrammatic evaluation we demand that  
at least four lines meet at every inner vertex,  
\begin{equation}  
x_{I_1,I_2|\vecm}= 0 \quad \hbox{for} \quad |I_1|+|I_2|=2 \; .  
\label{xi1i2eq2} 
\end{equation}  
We can fulfill these conditions by restricting our parameters  
$ \lambda_{I_1,I_2|\vecm}$. Using the form~(\ref{PGsqnew}) together  
with~(\ref{xi1i2eq2}) one easily sees that   
\begin{eqnarray}  
\langle\hat{P}_{\rm G|\vecm}^2\rangle_0 &=& 1 \; ,  
\label{one}  
\\  
\langle \hat{c}_{\vecm,\sigma}^+  
 \hat{P}_{\rm G|\vecm}^2  
\hat{c}_{\vecm,\sigma'} \rangle_0  
&=& \langle \hat{c}_{\vecm,\sigma}^+ \hat{c}_{\vecm,\sigma'}   
\rangle_0 \; ,  
\label{two}  
\\  
\langle\hat{c}_{\vecm,\sigma}^+  
 \hat{P}_{\rm G|\vecm}^2   
 \hat{c}_{\vecm,\sigma'}^+ \rangle_0  
&=& \langle \hat{c}_{\vecm,\sigma}^+ \hat{c}_{\vecm,\sigma'}^+   
\rangle_0 \; ,  
\label{three}  
\\  
\langle \hat{c}_{\vecm,\sigma}  
\hat{P}_{\rm G|\vecm}^2 
\hat{c}_{\vecm,\sigma'} \rangle_0  
&=& \langle \hat{c}_{\vecm,\sigma} \hat{c}_{\vecm,\sigma'}   
\rangle_0  
\label{four}  
\end{eqnarray}  
must be fulfilled. This follows from the fact that, apart from  
the trivial term in~(\ref{PGsqnew}), at least  
four lines meet at every vertex; a contraction with two `external'  
fermionic operators leaves at least one uncontracted pair whose local  
contraction, however, vanishes by construction of new local  
operators~(\ref{PGsqnew}). Explicitly,  
\begin{eqnarray}  
\sum_{I_1,I_2} \overline{\lambda}_{I_1,I_2|\vecm}  
m_{I_1,I_2|\vecm}^0 
&=& 1 \; , \label{neb1} 
\\  
\sum_{I_1(\sigma \in I_1)}\sum_{I_2(\sigma' \in I_2)} 
\overline{\lambda}_{I_1\setminus \sigma,I_2\setminus \sigma'|\vecm} 
m_{I_1,I_2|\vecm}^0 
&=& \langle \hat{c}_{\vecm,\sigma}^+ \hat{c}_{\vecm,\sigma'}   
\rangle_0 \; ,  
\label{neb2}
\\  
\sum_{I_1(\sigma \in I_1)}\sum_{I_2(\sigma' \in I_2)} 
\overline{\lambda}_{I_1\setminus \sigma,I_2|\vecm} 
m_{I_1,I_2\setminus \sigma'|\vecm}^0 
&=& \langle \hat{c}_{\vecm,\sigma}^+ \hat{c}_{\vecm,\sigma'}^+  
\rangle_0 \; ,  
\label{neb3}
\\  
\sum_{I_1(\sigma \in I_1)}\sum_{I_2(\sigma' \in I_2)} 
\overline{\lambda}_{I_1,I_2 \setminus \sigma'|\vecm} 
m_{I_1\setminus \sigma,I_2|\vecm}^0 
&=& \langle \hat{c}_{\vecm,\sigma} \hat{c}_{\vecm,\sigma'}   
\rangle_0 \; .  \label{neb4} 
\end{eqnarray}  
Here, we introduced the expectation value 
\begin{eqnarray} 
 m_{I_1,I_2|\vecm}^0&=&\langle \Psi_0 | I_1\rangle_{\vecm}   
{}_{\vecm}\langle I_2| \Psi_0\rangle \nonumber \\ 
&=&\langle \Psi_0 | 
\hat{n}_{I_1,I_2|\vecm}\hat{n}^{\rm h}_{I_1\cup I_2|\vecm} 
| \Psi_0\rangle \; , 
\end{eqnarray}  
with 
\begin{equation} 
\hat{n}^{\rm h}_{I|\vecm}\equiv  
\prod_{\sigma \in \overline{I}}(1-\hat{n}_{\vecm,\sigma}) \;  
\end{equation}  
and 
\begin{equation}
\hat{n}_{\vecm,\sigma}\equiv\hat{c}_{\vecm,\sigma}^+ \hat{c}_{\vecm,\sigma} \;.
\end{equation}   
Equations (\ref{neb1})--(\ref{neb4})  
can be used to fix some of the variational  
parameters $\lambda_{I_1,I_2|\vecm}$, for example those with  
$|I_1|+|I_2|\leq 1$.  
  
\subsection{Diagrammatics}  
\label{subsec:diagrammatics}
  
With the help of~(\ref{PGsqnew}) we expand the product over   
the squares of local Gutzwiller 
correlators in~(\ref{Oi}) [and in~(\ref{Oij})]  
in the form  
\begin{equation}  
\prod_{\vecm\neq \veci,\vecj}\hat{P}_{\rm G;\vecm}^2  
= 1+ \sumprime_{\vecm}\left[\hat{P}_{\rm G;\vecm}^2\right]^{\rm HF}   
+\frac{1}{2}\sumprime_{\vecm_1,\vecm_2}  
\left[\hat{P}_{\rm G;\vecm_1}^2\right]^{\rm HF}   
\left[\hat{P}_{\rm G;\vecm_2}^2\right]^{\rm HF} +\ldots\; ,  
\end{equation}  
where the prime on the sums indicates that all lattice sites  
are different from each other and from $\veci$ [and $\vecj$].  
When Wick's theorem is applied to the numerators in~(\ref{Oi})   
and in~(\ref{Oij}) we can introduce {\em new contractions\/}  
between two Fermi operators $\hat{a}_{\vecm,\sigma}$ and  
$\hat{a}_{\vecn,\sigma}$ ($\hat{a}=\hat{c}$ or $\hat{a}=\hat{c}^+$)  
\begin{equation}  
K_{\vecm,\sigma;\vecn,\sigma'}^0 =  
\langle \hat{a}_{\vecn,\sigma}\hat{a}_{\vecm,\sigma'}\rangle_0  
-\delta_{\vecn,\vecm}   
\langle \hat{a}_{\vecn,\sigma}\hat{a}_{\vecn,\sigma'}\rangle_0 \; .  
\label{defnewK}  
\end{equation}  
In most cases, $\vecn\neq \vecm$ holds and the new definition   
of a contraction reduces  
to the usual one because the extra term proportional to  
$\delta_{\vecn,\vecm}$ vanishes. In those cases where   
Fermi operators on the same site are contracted the contribution  
must vanish because we subtracted all local Hartree contributions in   
$\left[\hat{P}_{\rm G;\vecm}^2\right]^{\rm HF}$. The new contraction   
in~(\ref{defnewK}) fulfills this condition.  
  
The application of Wick's theorem thus gives a result which 
we would have obtained if we had worked with Grassmann operators  
instead of Fermion operators in the definition of the expectation values; 
all Grassmann operators~$\hat{a}^g_{\veci,\sigma}$ 
anti-commute with each other
\begin{equation}
\left[ \hat{a}^g_{\veci,\sigma}, \hat{a}^g_{\vecj,\sigma'}\right]_+ =0 \; ,
\end{equation} 
so that local contractions always vanish,
\begin{equation}
\langle \hat{a}^g_{\veci,\sigma} \hat{a}^g_{\veci,\sigma'}\rangle_0 =0 \; .
\end{equation} 
The use of Grassmann operators  
instead of Fermi operators also shows that we may now drop  
the restrictions on the lattice sums because all contributions  
with two lattice sites put equal vanish due to the anti-commutation  
relation between the corresponding Grassmann operators.  
In this way we have generated a diagrammatic theory in which   
lines between two vertices~$\vecn$ and $\vecm$  
are given by the one-particle  
density matrices $K_{\vecm,\sigma;\vecn,\sigma'}^0$  
defined in~(\ref{defnewK}), and $x_{I_1,I_2|\vecm}$ 
gives the strength of a vertex~$(I_1,I_2|\vecm)$  
with $|I_1|$ in-going lines and $|I_2|$ out-going lines. 
 
Now we are in the position to apply the linked-cluster 
theorem so that all disconnected diagrams in the numerator  
in~(\ref{Oi}) [and in~(\ref{Oij})] cancel the corresponding   
denominator. Then, the calculation of the 
expectation values~(\ref{Oi}) and~(\ref{Oij})  
is reduced to the sum over all connected diagrams 
according to the Feynman rules with lines and vertices as defined 
above. 
 
\section{Exact results for infinite coordination number}
\label{exactdinfty}
\label{sect:calcs}  
 
By construction, at least four lines meet at every inner  
vertex. Due to the absence of local Hartree--Fock contributions  
two inner vertices are always connected by at least three  
independent paths of lines (there is no diagram with a single inner  
vertex). For lattices with $Z$ nearest neighbors, 
\begin{equation}  
K_{\vecm,\sigma|\vecn,\sigma'}^0 \sim {\cal O}\left(  
Z^{-|\vecn-\vecm|/2}  
\right) \; ; 
\end{equation}  
for a simple cubic lattice in $d$~dimensions, $Z=2d$. 
Therefore, the contribution of all diagrams with inner vertices vanishes  
at least proportional to $1/\sqrt{d}$ in the limit of infinite  
dimensions. Thus, not a single diagram needs to  
be calculated in infinite dimensions, i.e., we arrive at the result  
\begin{eqnarray}  
\langle \hat{O}_{\veci} \rangle_{\rm G} &=&  
 \langle \Psi_0 |   
\hat{P}_{\rm G|\veci}\hat{O}_{\veci}\hat{P}_{\rm G|\veci} |  
\Psi_0\rangle \; ,  
\label{Oired}  
\\  
\langle \hat{O}_{\veci, \vecj} \rangle_{\rm G} &=&  
\langle \Psi_0 |   
\hat{P}_{\rm G|\veci}\hat{P}_{\rm G|\vecj}  
\hat{O}_{\veci,\vecj}\hat{P}_{\rm G|\veci} \hat{P}_{\rm G|\vecj}  
|\Psi_0\rangle   
\label{Oijred}  
\end{eqnarray}  
in $d=\infty$ dimensions. The result~(\ref{Oijred}) can be further  
simplified because in $d=\infty$ dimensions there can be  
only a single line connecting the two external vertices~$\veci$  
and~$\vecj$.  
  
\subsection{Local quantities}  
\label{subsec:localquantities}
 
Now we address the local operators for the particle densities,  
\begin{equation}  
\hat{n}_{\veci,\sigma \sigma'} \equiv
\hat{c}_{\veci,\sigma}^+\hat{c}_{\veci,\sigma'}\; ,  
\end{equation}  
the pairing densities,  
\begin{equation}  
\hat{s}_{\veci,\sigma \sigma'}\equiv
\hat{c}_{\veci,\sigma}\hat{c}_{\veci,\sigma'}\; ,  
\end{equation}  
and the local interaction  
\begin{equation}  
\hat{H}_{\rm at}\equiv  \sum_{\vecm} \sum_{J_1,J_2} U_{J_1,J_2|\vecm}  
|J_1\rangle_{\vecm}   
{}_{\vecm}\langle J_2| \; .  
\label{Hint}  
\end{equation}  
In infinite dimensions we have   
\begin{eqnarray}  
n_{\veci,\sigma \sigma'} &=& \langle \hat{n}_{\veci,\sigma \sigma'}  
\rangle_{\rm G} \nonumber= \langle  \hat{P}_{\rm G|\veci}  
\hat{c}_{\veci,\sigma}^+\hat{c}_{\veci,\sigma'}  
 \hat{P}_{\rm G|\veci}\rangle_0 \nonumber  \\  
&=& \sum_{I_1,I_4} N_{I_1,I_4|\veci}^{\sigma\sigma'}  
m_{I_1,I_4|\veci}^0  \; .
\end{eqnarray}  
Here we defined the matrix elements for the local densities as  
\begin{eqnarray}  \nonumber  
 N_{I_1,I_4|\veci}^{\sigma\sigma'}   &\equiv& \sum_{I_2,I_3}  
 \lambda_{I_1,I_2|\veci} \lambda_{I_3,I_4|\veci} \,   
 {}_{\veci}\langle I_2|\hat{c}_{\veci,\sigma}^+\hat{c}_{\veci,\sigma'}  
 | I_3\rangle_{\veci} \; \\ 
&=&\sum_{I(\sigma \sigma'\notin I)} 
\lambda_{I_1,I \cup \sigma|\veci} \lambda_{I\cup \sigma',I_4|\veci}  
\fsgn(\sigma,I)\fsgn(\sigma',I)\;   
\end{eqnarray}  
where  
\begin{equation} 
\fsgn(\sigma,I)\equiv\langle I\cup \sigma|\hat{c}_{\sigma}^{+}|I \rangle 
\end{equation} 
is $\pm1$ if it takes an odd/even number of anti-commutations to shift the  
state $\sigma$ to its proper place in the ordered sequence~$I$. 
Correspondingly, we find  
\begin{eqnarray}  
s_{\veci,\sigma \sigma'} &=& \langle \hat{s}_{\veci,\sigma \sigma'}  
\rangle_{\rm G} \nonumber = \langle  \hat{P}_{\rm G|\veci}  
\hat{c}_{\veci,\sigma}\hat{c}_{\veci,\sigma'}  
 \hat{P}_{\rm G|\veci}\rangle_0 \\  
&=& \sum_{I_1,I_4} S_{I_1,I_4|\veci}^{\sigma\sigma'}  
m_{I_1,I_4|\veci}^0 \; .  
\end{eqnarray}  
Here we defined the matrix elements for the local pairing amplitudes as  
\begin{eqnarray}  \nonumber  
 S_{I_1,I_4|\veci}^{\sigma\sigma'}  &\equiv& \sum_{I_2,I_3}  
 \lambda_{I_1,I_2|\veci} \lambda_{I_3,I_4|\veci}\,   
 {}_{\veci}\langle I_2|\hat{c}_{\veci,\sigma}\hat{c}_{\veci,\sigma'}  
 | I_3\rangle_{\veci} \\ 
&=&\sum_{I(\sigma \sigma'\notin I)} 
\lambda_{I_1,I|\veci} \lambda_{I\cup (\sigma \sigma'),I_4|\veci} 
\fsgn(\sigma,I)\fsgn(\sigma',I) \fsgn(\sigma,\sigma') \; . 
\end{eqnarray}  
Lastly, along the same lines we find   
\begin{eqnarray}  
E^{\rm at} &=& \langle \hat{H}_{\rm at}  
\rangle_{\rm G} \nonumber \\ 
&=&  \sum_{\veci}\langle  
 \hat{P}_{\rm G|\veci} \sum_{J_1,J_2}  
U_{J_1,J_2|\veci} |J_1\rangle_{\veci} {}_{\veci}\langle J_2|  
 \hat{P}_{\rm G|\veci}\rangle_0 \label{Eint}\\   
&=& \sum_{\veci} \sum_{I_1,I_4} \overline{U}_{I_1,I_4|\veci}  
  m_{I_1,I_4|\veci}^0   
\; ,   \nonumber 
\end{eqnarray}  
where   
\begin{equation}  
 \overline{U}_{I_1,I_4|\veci} \equiv \sum_{I_2,I_3}  
 \lambda_{I_1,I_2|\veci} \lambda_{I_3,I_4|\veci}U_{I_2,I_3|\veci}  
  \; .   
\label{Utilde}  
\end{equation}  
 
\subsection{Single-particle density matrices}  
\label{subsec:singlep}
  
Now we evaluate the single-particle density matrix  
\begin{equation}  
P_{\veci,\sigma;\vecj\sigma'} \equiv \langle    
\hat{c}_{\veci,\sigma}^+\hat{c}_{\vecj,\sigma'}  
\rangle_{\rm G} \; 
\end{equation}  
and the one-particle pairing matrix  
\begin{equation}  
S_{\veci,\sigma;\vecj\sigma'} \equiv \langle    
\hat{c}_{\veci,\sigma}\hat{c}_{\vecj,\sigma'}  
\rangle_{\rm G}   
\end{equation}  
for $\veci\neq\vecj$.

 We start with $\hat{O}_{\veci\vecj}=  
\hat{c}_{\veci,\sigma}^+\hat{c}_{\vecj,\sigma'}$ in~(\ref{Oij})  
and evaluate  
\begin{equation}  
\hat{P}_{\rm G|\veci}\hat{c}_{\veci,\sigma}^+  
\hat{P}_{\rm G|\veci}  
=  \sum_{I_1,I_4} c_{I_1,I_4|\veci,\sigma}^*  
 | I_1\rangle_{\veci} {}_{\veci}\langle I_4|   
\end{equation}  
with  
\begin{eqnarray}  
c_{I_1,I_4|\veci,\sigma}^*&\equiv&   
 \sum_{I_2,I_3}  
 \lambda_{I_1,I_2|\veci} \lambda_{I_3,I_4|\veci}\,   
 {}_{\veci}\langle I_2|\hat{c}_{\veci,\sigma}^+ | I_3\rangle_{\veci} 
\nonumber \\ 
&=& \sum_{I(\sigma \notin I)}\fsgn(\sigma,I)\lambda_{I_1,I\cup\sigma|\veci} 
 \lambda_{I,I_4|\veci}\; .  
\end{eqnarray}  
In $d=\infty$ dimensions only a single line can join the two external  
vertices $\veci\neq\vecj$ in~(\ref{Oijred}). In the first  
case, the contraction at~$\veci$ is done with  
a creation operator~$\hat{c}_{\veci,\gamma}^+$. There are 
two possibilities, it can either be $\gamma \in I_1$, with 
\begin{equation} 
| I_1\rangle_{\veci} {}_{\veci}\langle I_4|= 
\left(\fsgn(\gamma,I_1\setminus \gamma)\, 
\hat{n}_{I_1\setminus \gamma,I_4|\veci }\, 
\hat{n}^{\rm h}_{I_1\cup I_4|\veci }\right) 
\hat{c}_{\veci,\gamma}^{+}\;, 
\end{equation} 
or  
$\gamma \in \overline{I_1 \cup I_4}$, with 
\begin{equation} 
| I_1\rangle_{\veci} {}_{\veci}\langle I_4|= 
-\left(\fsgn(\gamma,I_4)\, 
\hat{n}_{I_1,I_4\cup \gamma|\veci }\, 
\hat{n}^{\rm h}_{I_1\cup I_4\cup \gamma|\veci }\right) 
\hat{c}_{\veci,\gamma}^{+}\;. 
\end{equation} 
Altogether, the amplitude for such a process is given by 
\begin{equation}  
q_{\veci,\sigma\gamma}^{*} \equiv 
\sum_{I(\gamma \notin I)} \sum_{I'}\fsgn(\gamma,I) 
(m_{I,I'|\veci}^{0(\gamma)}c_{I\cup\gamma,I'|\veci,\sigma}^* 
-m_{I',I\cup\gamma|\veci}^{0}c_{I',I|\veci,\sigma}^*) \; .  
\label{Qvalues}  
\end{equation}  
Here, we introduced the expectation value 
\begin{equation} 
m_{I,I'|\veci}^{0(\gamma)}\equiv\langle 
\hat{n}_{I,I'|\veci }\, 
\hat{n}^{\rm h}_{I\cup I'\cup \gamma|\veci } 
\rangle_{0}\;. 
\end{equation}  
In previous works~\cite{PRB,EPL},  
the factors $q_{\veci,\sigma\gamma}^{*}$ were denoted by  
$\sqrt{q_{\veci,\sigma\gamma}}$. In general, however,
$q_{\veci,\sigma\gamma}^{*}$ is 
not a positive real number.  
 
The second case is only possible if there are superconducting pairing  
correlations in $|\Psi_0\rangle$. Then, the line between $\veci$ and $\vecj$  
can also be an anomalous propagator, i.e., the contraction at~$\veci$  
can be done with an annihilation operator $\hat{c}_{\veci,\gamma}$. 
Again, we obtain two contributions,  $\gamma \in I_4$, which gives 
\begin{equation} 
| I_1\rangle_{\veci} {}_{\veci}\langle I_4|= 
\hat{c}_{\veci,\gamma}^{} 
\left(\fsgn(\gamma,I_2\setminus \gamma)\, 
\hat{n}_{I_1,I_4\setminus \gamma|\veci }\, 
\hat{n}^{\rm h}_{I_1\cup I_4|\veci }\right)\;, 
\end{equation} 
 and 
$\gamma \in \overline{I_1 \cup I_4}$, which contributes 
\begin{equation} 
| I_1\rangle_{\veci} {}_{\veci}\langle I_4|= 
\hat{c}_{\veci,\gamma}^{} 
\left(\fsgn(\gamma,I_1)\, 
\hat{n}_{I_1\cup \gamma,I_4|\veci }\, 
\hat{n}^{\rm h}_{I_1\cup I_4\cup \gamma|\veci }\right)\;. 
\end{equation} 
Therefore, the amplitude for the anomalous process can be written as 
\begin{equation}  
\bar{q}_{\veci,\sigma\gamma}^{} \equiv \sum_{I(\gamma \notin I)} \sum_{I'}\fsgn(\gamma,I) 
(m_{I',I|\veci}^{0(\gamma)}c_{I',I\cup\gamma|\veci,\sigma}^* 
+m_{I\cup\gamma,I'|\veci}^{0}c_{I,I'|\veci,\sigma}^*) 
\; .  
\label{Avalues}  
\end{equation}  
With these definitions we find for the one-particle density matrix 
\begin{eqnarray}  
P_{\veci,\sigma;\vecj\sigma'} &=& \sum_{\gamma \gamma'} \Bigl[  
q_{\veci,\sigma\gamma}^{*}
 q^{}_{\vecj,\sigma'\gamma'}   
\langle  \hat{c}_{\veci,\gamma}^+\hat{c}_{\vecj,\gamma'}\rangle_0  
+ \bar{q}_{\veci,\sigma\gamma}^{} \bar{q}_{\vecj,\sigma'\gamma'}^{*}  
\langle  \hat{c}_{\veci,\gamma}\hat{c}_{\vecj,\gamma'}^+\rangle_0  
\nonumber \\  
&&\hphantom{\sum_{\gamma \gamma'}l}  
q_{\veci,\sigma\gamma}^{*}
 \bar{q}_{\vecj,\sigma'\gamma'}^{*}   
\langle  \hat{c}_{\veci,\gamma}^+\hat{c}_{\vecj,\gamma'}^+\rangle_0  
+ \bar{q}_{\veci,\sigma\gamma}^{} q^{}_{\vecj,\sigma'\gamma'}  
\langle  \hat{c}_{\veci,\gamma}\hat{c}_{\vecj,\gamma'}\rangle_0  
\Bigr] 
\end{eqnarray}  
in the limit of infinite dimensions.  
 
Analogously, for the one-particle pairing matrix we find  
\begin{eqnarray}  
S_{\veci,\sigma;\vecj\sigma'} &=& \sum_{\gamma \gamma'} \Bigl[  
q_{\veci,\sigma\gamma}^{} q^{}_{\vecj,\sigma'\gamma'}   
\langle  \hat{c}_{\veci,\gamma}\hat{c}_{\vecj,\gamma'}\rangle_0  
+ \bar{q}_{\veci,\sigma\gamma}^{*} \bar{q}_{\vecj,\sigma'\gamma'}^{*}  
\langle  \hat{c}_{\veci,\gamma}^+\hat{c}_{\vecj,\gamma'}^+\rangle_0  
\nonumber \\  
&&\hphantom{\sum_{\gamma \gamma'}l}  
q_{\veci,\sigma\gamma}^{} \bar{q}_{\vecj,\sigma'\gamma'}^{*}   
\langle  \hat{c}_{\veci,\gamma}\hat{c}_{\vecj,\gamma'}^+\rangle_0  
+ \bar{q}_{\veci,\sigma\gamma}^{*} q^{}_{\vecj,\sigma'\gamma'}  
\langle  \hat{c}_{\veci,\gamma}^+\hat{c}_{\vecj,\gamma'}\rangle_0  
\Bigr] 
\end{eqnarray}  
in the limit of infinite dimensions.

\section{Variational ground-state energy}  
\label{gsenergy}  
\label{groundstate}
 
In this section we specify the variational problem in the limit of 
high dimensions. 
We use translational invariance to render the problem 
numerically tractable.

\subsection{Variational problem}  
\label{subsec:Variational}

The variational ground-state energy for the multi-band Hamiltonian
(\ref{hamiltonian5}) is given by  
\begin{equation}  
E^{\rm var}(\{\lambda_{I_1,I_2|\veci}\},\{|\Psi_0\rangle\})  
= \langle \hat{H}\rangle_{\rm G} \;.  
\end{equation}  
With the help of the results of Section~\ref{sect:calcs},  
this expectation value is readily evaluated in infinite dimensions.  
We find  
\begin{eqnarray} \label{fullEvar} 
E^{\rm var}(\{\lambda_{I_1,I_2|\veci}\},\{|\Psi_0\rangle\})  
&=&  
  E^{\rm kin}(\{\lambda_{I_1,I_2|\veci}\},\{|\Psi_0\rangle\})  
+ E^{\rm at}(\{\lambda_{I_1,I_2|\veci}\},\{|\Psi_0\rangle\}) 
\; ,  \\[5pt]  \label{Ekin}  
E^{\rm kin}(\{\lambda_{I_1,I_2|\veci}\},\{|\Psi_0\rangle\})  
&=& \sum_{\veci,\vecj;\sigma \sigma'}\sum_{\alpha, \alpha'=+,-}   
t_{\veci,\vecj}^{\alpha \alpha';\sigma \sigma'}   
\langle \Psi_0 |\hat{c}_{\veci,\sigma}^{\alpha}
\hat{c}_{\vecj,\sigma'}^{\alpha'}  
|\Psi_0\rangle
\; ,\\[5pt]
E^{\rm at}(\{\lambda_{I_1,I_2|\veci}\},\{|\Psi_0\rangle\})  
&=&  \sum_{\veci} \sum_{I_1,I_4} \overline{U}_{I_1,I_4|\veci}  
  m_{I_1,I_4|\veci}^0   \; .      
\end{eqnarray}  
where $\overline{U}_{I_1,I_4|\veci}$ is given in~(\ref{Utilde}).
The superscripts $\alpha, \alpha'=+/-$ were introduced in order to distinguish
creation and annihilation operators ($\hat{c}^{-}\equiv\hat{c}^{}$). 
Mathematically, these superscripts are interpreted as real 
numbers, e.g., $\hat{c}^{(-\alpha)}$ is equal to $\hat{c}^{+}$ or
$\hat{c}$ when $\alpha=-$ or $\alpha=+$, respectively. 
For the hopping amplitudes in (\ref{Ekin}) we find
\begin{eqnarray}  
t_{\veci,\vecj}^{+-;\sigma \sigma'} &=&\frac{1}{2} \sum_{\tau \tau'}  
 \left(q_{\veci,\tau\sigma}^{*}
 q^{}_{\vecj,\tau'\sigma'}t_{\veci,\vecj}^{\tau \tau'} 
-\bar{q}_{\veci,\tau\sigma}^{*} \bar{q}_{\vecj,\tau'\sigma'}^{}
t_{\vecj,\veci}^{\tau',\tau}\right)  \; ,\; 
t_{\veci,\vecj}^{-+;\sigma \sigma'}=
-\left(t_{\veci,\vecj}^{+-;\sigma \sigma'}\right)^*\;,\\
t_{\veci,\vecj}^{++;\sigma \sigma'} &=& \frac{1}{2} \sum_{\tau \tau'}  
 \left(q_{\veci,\tau\sigma}^{*}
\bar{q}_{\vecj,\tau'\sigma'}^{*}
t_{\veci,\vecj}^{\tau \tau'} 
-\bar{q}_{\veci,\tau\sigma}^{*} 
q_{\vecj,\tau'\sigma'}^{*} 
t_{\vecj,\veci}^{\tau',\tau}\right)  \; ,\; 
t_{\veci,\vecj}^{--;\sigma \sigma'}=
-\left(t_{\veci,\vecj}^{++;\sigma \sigma'}\right)^*\;.
\end{eqnarray}  
The interaction energy is given in terms of the variational   
parameters~$\lambda_{I_1,I_2|\veci}$ and purely local expectation  
values in $|\Psi_0\rangle$. With the help of Wick's theorem  
these expectation values can be expressed solely in terms of the local  
single-particle density matrix $\mymatrix{C}_{\vecm}$ with entries 
\begin{equation}  
C_{\vecm,\sigma\sigma'}^{\alpha \alpha'} 
=\langle \hat{c}_{\vecm,\sigma}^{\alpha} 
 \hat{c}_{\vecm,\sigma'}^{\alpha'}   
\rangle_0 \;.  \label{cmat} 
\end{equation}

\subsection{Minimization}  
\label{subsec:mini}
 
\subsubsection{Effective single-particle Schr\"odinger equation}  

We have to optimize the variational ground-state energy~(\ref{fullEvar})   
with respect to all parameters $\{\lambda_{I_1,I_2|\veci}\}$ and the  
set of all one-particle quasi-particle vacua $\{|\Psi_0\rangle\}$,  
whereby the equations~(\ref{neb1})--(\ref{neb4}) have to be  
obeyed. The latter may be fulfilled by fixing the parameters  
$\{\lambda_{I_1,I_2|\veci}\}$ for $|I_1|,|I_2|\leq 1$,  
which are then functions of $\mymatrix{C}_{\veci}$ and  
$\{\mymatrix{\lambda}_{\veci}\}\equiv\{\lambda_{I_1,I_2|\veci}\}$  
with $|I_1|,|I_2|\geq 2$.  
We introduce the tensors $\mymatrix{\eta}_{\veci}$ of Lagrange multipliers
 with  entries $\eta_{\veci,\sigma\sigma'}^{\alpha,\alpha'}$ and 
the multiplier  $E^{\rm SP}$ in order to fulfill equation~(\ref{cmat}) 
and to ensure the normalization of  $| \Psi_0 \rangle$. 
In addition, we fix the average number of particles $N=nL$ in 
$|\Psi_{\rm G}\rangle$ with the help
 of the Lagrange multiplier~$\Lambda$.  
Then, the variational  
ground-state energy $E_0^{\rm var}$ is given by  
\begin{eqnarray}  
E_0^{\rm var}&=& 
\mathop{\rm Minimum}_{\{\mymatrix{\lambda}_{\veci}\},\{\mymatrix{C}_{\veci}\}  
,\{\mymatrix{\eta}_{\veci}\},\Lambda,\{|\Psi_0\rangle\}
,E^{\rm SP}}  
E_{\rm c}\left(\{\mymatrix{\lambda}_{\veci}\},\{\mymatrix{C}_{\veci}\}  
,\{\mymatrix{\eta}_{\veci}\},\Lambda,\{|\Psi_0\rangle\}
,E^{\rm SP}\right) \\[5pt]
E_{\rm c}(\ldots) 
&=&  
E^{\rm var}\left(\{\mymatrix{\lambda}_{\veci}\},\{\mymatrix{C}_{\veci}\} 
\{|\Psi_0\rangle\}\right)   
-E^{\rm SP}\left(\langle \Psi_0 |\Psi_0\rangle -1\right)  
\nonumber \\  
&& -\sum_{\veci,\sigma\sigma'}  \sum_{\alpha,\alpha'}
\left[ \eta_{\veci,\sigma\sigma'}^{\alpha\alpha'}
\left(C^{\alpha\alpha'}_{\veci,\sigma\sigma'}-\langle \Psi_0 |
 \hat{c}_{\veci,\sigma}^{\alpha}   
\hat{c}_{\veci,\sigma'}^{\alpha'}  | \Psi_0 \rangle \right) 
+{\rm c.c.}  \right]
\label{bigminimization}  
\\
&&+\Lambda \biggl(\sum_{\veci,\sigma} 
n_{\veci,\sigma\sigma}\left(\{\mymatrix{\lambda}_{\veci}\},
\{\mymatrix{C}_{\veci}\}\right)-N\biggr)\; .  
 \nonumber   
\end{eqnarray}  
The minimization with respect to~$|\Psi_0\rangle$ can be   
carried out explicitly and leads to the effective  
one-particle Schr\"odinger equation  
\begin{eqnarray}
 \hat{T}^{\rm eff}| \Psi_0 \rangle  
&=&E^{\rm SP}\left(\{\mymatrix{\lambda}_{\veci}\}
,\{\mymatrix{C}_{\veci}\} ,\{\mymatrix{\eta}_{\veci}\}\right) 
|\Psi_0\rangle \; , \\
\hat{T}^{\rm eff}&\equiv&\frac{1}{2}\sum_{\veci,\sigma;\vecj,\sigma'}
\sum_{\alpha \alpha'}\left[ \left(  
t_{\veci,\vecj}^{\alpha \alpha';\sigma \sigma'}   
+ 2\delta_{\veci,\vecj}
\eta^{\alpha \alpha'}_{\veci,\sigma\sigma'}\right)  
\hat{c}_{\veci,\sigma}^{\alpha}   
\hat{c}_{\vecj,\sigma'}^{\alpha'} +{\rm h.c.} \right] 
\nonumber \\
&=&T_0+\frac{1}{2}\sum_{\veci,\sigma;\vecj,\sigma'}
\sum_{\alpha \alpha'}\left[\left[  
t_{\veci,\vecj}^{\alpha \alpha';\sigma \sigma'}   
+ \delta_{\veci,\vecj}
\left(\eta^{\alpha \alpha'}_{\veci,\sigma\sigma'}
-\eta^{\alpha' \alpha}_{\veci,\sigma'\sigma}\right)
\right]  
\hat{c}_{\veci,\sigma}^{\alpha}   
\hat{c}_{\vecj,\sigma'}^{\alpha'} +{\rm h.c.}\right]
\;,
\label{SPSEq}  
\end{eqnarray} 
where 
\begin{equation}
T_0=\sum_{\veci,\sigma}\left[
\left(\eta^{+-}_{\veci,\sigma \sigma}+\eta^{-+}_{\veci,\sigma \sigma}
\right)
+{\rm c.c.}\right]\;.
\end{equation} 
We assume that the optimum wave-function $|\Psi_0\rangle$ is the 
ground-state of the Hamiltonian~(\ref{SPSEq}). 
In this way, $|\Psi_0\rangle$ becomes a function of  
$\{\mymatrix{\lambda}_{\veci}\},\{\mymatrix{C}_{\veci}\}  
,\{\mymatrix{\eta}_{\veci}\}$,  
 and the remaining task is to find the minimum
\begin{eqnarray}  
E_0^{\rm var}&=& \mathop{\rm Minimum}_{\{\mymatrix{\lambda}_{\veci}\}, 
\{\mymatrix{C}_{\veci}\}  
,\{\mymatrix{\eta}_{\veci}\},\Lambda} 
E_{\rm c}(\{\mymatrix{\lambda}_{\veci}\}, 
\{\mymatrix{C}_{\veci}\}  
,\{\mymatrix{\eta}_{\veci}\},\Lambda)\;, \\[3pt]
E_{\rm c}(\ldots) 
&=&  
E^{\rm SP}  
\left(\{\mymatrix{\lambda}_{\veci}\},\{\mymatrix{C}_{\veci}\}  
,\{\mymatrix{\eta}_{\veci}\}\right)  
+ E^{\rm at} \left(\{\mymatrix{\lambda}_{\veci}\},\{\mymatrix{C}_{\veci}\} 
\right)  \\ \nonumber 
&&  
- \sum_{\veci,\sigma\sigma'}\left[\sum_{\alpha \alpha'}  
\left(\eta_{\veci,\sigma\sigma'}^{\alpha \alpha'}
  C_{\veci,\sigma\sigma'}^{\alpha \alpha'}+{\rm c.c.}\right)
-
\delta_{\sigma\sigma'}
\Lambda\left(n_{\veci,\sigma\sigma}\left(\{\mymatrix{\lambda}_{\veci}\}, 
\{\mymatrix{C}_{\veci}\}
\right)
-n\right)  
\right] \; .  
\label{Minimizeit}  
\end{eqnarray}

\subsubsection{Translational invariance}  
 
To make further progress we assume translational invariance.  
Then, the single-particle Hamiltonian in the Bloch basis with 
wave vectors~$\veck$ has the form
\begin{equation}  
\hat{T}^{\rm eff} =T_0+ \sum_{\veck;\sigma \sigma'}\sum_{\alpha \alpha'}
\epsilon^{\alpha \alpha'}_{\sigma \sigma'}(\veck)
\hat{c}_{(\alpha\veck),\sigma}^{\alpha}   
\hat{c}_{(-\alpha'\veck),\sigma'}^{\alpha'}\;,
\label{Teff}  
\end{equation}  
where the coefficients
\begin{eqnarray} \nonumber
\epsilon^{\alpha \alpha'}_{\sigma \sigma'}(\veck)&=&
\frac{1}{L}\sum_{\veci,\vecj}e^{-\mathi\veck (\vecj-\veci)}
\left(t_{\veci,\vecj}^{\alpha \alpha';\sigma \sigma'}   
+ \frac{1}{2}\delta_{\veci,\vecj}   
\left(\eta^{\alpha \alpha'}_{\sigma\sigma'}-
\eta^{\alpha'\alpha}_{\sigma'\sigma}+
\left[\eta^{(-\alpha') (-\alpha)}_{\sigma'\sigma}\right]^{*}-
\left[\eta^{(-\alpha)(-\alpha')}_{\sigma\sigma'}\right]^{*}
\right)
\right)\\
&=&\frac{1}{2}\sum_{\tau \tau'}
Q_{\tau \tau'}^{\alpha \alpha';\sigma \sigma'}\epsilon^{0}_{\tau \tau'}(\veck)
+ \frac{1}{2}\left(
\eta^{\alpha \alpha'}_{\sigma\sigma'}-\eta^{\alpha'\alpha}_{\sigma'\sigma}
+\left[\eta^{(-\alpha')(-\alpha)}_{\sigma'\sigma}\right]^{*}-
\left[\eta^{(-\alpha)(-\alpha')}_{\sigma\sigma'}\right]^{*}
\right)
\label{epss}
\end{eqnarray} 
are the elements of the matrices 
$\widetilde{\epsilon}^{\alpha \alpha'}_{\veck}$.
In (\ref{epss}) we used the bare energy-band matrix (\ref{bareen}) and 
introduced the coefficients 
\begin{eqnarray}
Q_{\tau \tau'}^{+-;\sigma \sigma'}&=&
q_{\tau\sigma}^{*}q^{}_{\tau'\sigma'}
-\bar{q}_{\tau\sigma}^{*} \bar{q}_{\tau'\sigma'}^{}
  \; ,\; Q_{\tau \tau'}^{-+;\sigma \sigma'}=
-\left(Q_{\tau \tau'}^{+-;\sigma \sigma'}
\right)^{*}\\
Q_{\tau \tau'}^{{++};\sigma \sigma'}&=&
q_{\tau\sigma}^{*}\bar{q}_{\tau'\sigma'}^{*}
-\bar{q}_{\tau\sigma}^{*} q_{\tau'\sigma'}^{*} 
  \; ,\; Q_{\tau \tau'}^{--;\sigma \sigma'}=
-\left(Q_{\tau \tau'}^{++;\sigma \sigma'}
\right)^{*}\;.
\end{eqnarray}  
Let us regard the operators $\hat{c}_{\veck,\sigma}^{\alpha}$ 
as elements of vectors $\hat{\vecc}_{\veck}^{\alpha}$.
Then, the effective Hamiltonian $\hat{T}^{\rm eff}$ can be written 
in terms of matrix products,
\begin{equation}\label{teffb}
\hat{T}^{\rm eff}=T_0 + \sum_{\veck}
\left(
\begin{array}{c}
\widehat{\vecc}_{\veck}^{+}\\
\widehat{\vecc}_{-\veck}^{-}\\
\end{array}
\right)^{\rm T}
\left(
\begin{array}{cc}
\widetilde{\epsilon}^{+-}_{\veck}&\widetilde{\epsilon}^{++}_{\veck}\\
\widetilde{\epsilon}^{--}_{\veck}&\widetilde{\epsilon}^{-+}_{\veck}\\
\end{array}
\right)
\left(
\begin{array}{c}
\widehat{\vecc}_{\veck}^{-}\\
\widehat{\vecc}_{-\veck}^{+}\\
\end{array}
\right)\;.
\end{equation}
The matrix in (\ref{teffb}) may be diagonalized by means of a 
Bogoliubov transformation
\begin{equation} \label{Bogoliubov}
\left(
\begin{array}{c}
\widehat{\vecc}_{\veck}^{-}\\
\widehat{\vecc}_{-\veck}^{+}\\
\end{array}
\right)=
\left(
\begin{array}{cc}
\widetilde{u}^{--}_{\veck}&\widetilde{u}^{-+}_{\veck}\\
\widetilde{u}^{+-}_{\veck}&\widetilde{u}^{++}_{\veck}\\
\end{array}
\right)
\left(
\begin{array}{c}
\widehat{\vech}_{\veck}^{-}\\
\widehat{\vech}_{-\veck}^{+}\\
\end{array}
\right) \; ,
\end{equation}
where we introduced new fermionic operators
$\widehat{\vech}^{\alpha}_{\veck}$ and matrices 
$\widetilde{u}^{\alpha \alpha'}_{\veck}$ with elements 
$\widehat{h}^{\alpha}_{\veck,\gamma}$ and  
$u^{\alpha \alpha'}_{\sigma\gamma}(\veck)$, respectively.
For reasons of consistency the matrices 
$\widetilde{u}^{\alpha \alpha'}_{\veck}$  obey the symmetries 
\begin{equation}\label{cond1}
\widetilde{u}^{++}_{-\veck}=\left(\widetilde{u}^{--}_{\veck}\right)^{\ast}\;\;,\;\;
\widetilde{u}^{+-}_{-\veck}=\left(\widetilde{u}^{-+}_{\veck}\right)^{\ast}\;.
\end{equation} 
The fermionic commutation rules of the new operators 
$\widehat{\vech}_{\veck}^{\alpha}$ are ensured when
the matrix in~(\ref{Bogoliubov}) is unitary. In~(\ref{teffb})--(\ref{cond1})
we have used the
standard notations for the transposition $\widetilde{M}^{\rm T}$, the  
complex conjugate $\widetilde{M}^{\ast}$, and the 
Hermitian conjugate $\widetilde{M}^{\dagger}$ of a matrix~$\widetilde{M}$.

After the diagonalization $\hat{T}^{\rm eff}$ becomes
\begin{eqnarray}
\label{Teffdiaga}
\hat{T}^{\rm eff}&=&T_0+\frac{1}{2}
\sum_{\veck}\left(E_{\veck,\gamma}
\widehat{h}^{+}_{\veck,\gamma}
\widehat{h}_{\veck,\gamma}-E_{\veck,\gamma}
\widehat{h}_{\veck,\gamma}
\widehat{h}^{+}_{\veck,\gamma}\right)\nonumber \\
&=&T_0+E_0+\sum_{\veck}E_{\veck,\gamma}
\widehat{h}^{+}_{\veck,\gamma}
\widehat{h}_{\veck,\gamma}\;,\label{Teffdiag}
\end{eqnarray}
where the real quantities 
$E_{\veck,\gamma}=\delta_{\gamma \gamma'}E_{\gamma \gamma'}(\veck)$ are the 
elements of the diagonal matrix 
\begin{equation}
\label{ek}
\widetilde{E}_{\veck}=2\sum_{\alpha \alpha'}
\left(\widetilde{u}^{(-\alpha)-}_{\veck}\right)^{\dagger}
\widetilde{\epsilon}^{\alpha \alpha'}_{\veck}
\widetilde{u}^{\alpha'-}_{\veck}\;,
\end{equation}
  and 
\begin{equation}
E_0=-\frac{1}{2}\sum_{\veck}{\rm Tr}\left(\widetilde{E}_{\veck}\right)\;.
\end{equation}
For the derivation of (\ref{Teffdiaga})--(\ref{ek}) we have used the following
symmetries of the matrices $\widetilde{\epsilon}^{\alpha \alpha'}_{\veck}$, 
\begin{eqnarray}
\widetilde{\epsilon}^{\alpha \alpha'}_{-\veck}&=&
-\left(\widetilde{\epsilon}^{\alpha',\alpha}_{\veck}\right)^{\rm T}=
-\left(\widetilde{\epsilon}^{\alpha',\alpha}_{\veck}\right)^{*}
\qquad(\alpha \neq \alpha')\;,
\\
\widetilde{\epsilon}^{\alpha,\alpha}_{-\veck}&=&
-\left(\widetilde{\epsilon}^{\alpha,\alpha}_{\veck}\right)^{\rm T}=
-\left(\widetilde{\epsilon}^{\alpha',\alpha'}_{\veck}\right)^{*}
\qquad (\alpha \neq \alpha')
\end{eqnarray}
which follow from~(\ref{epss}) and
\begin{equation}
\epsilon^{0}_{\gamma',\gamma}(\veck)=
\left(\epsilon^{0}_{\gamma,\gamma'}(\veck)\right)^{*}
=\epsilon^{0}_{\gamma,\gamma'}(-\veck)
\end{equation} 
which result from hermiticity of $\hat{T}$ and the fact that our 
electron-transfer amplitudes are real.
Note that the matrices $\widetilde{u}^{\alpha,\alpha'}_{\veck}$, 
$\widetilde{\epsilon}^{\alpha \alpha'}_{\veck}$,
the operators $\widehat{h}_{\veck, \gamma }^{\alpha}$,
and the energies $E_{\veck,\gamma}$  
still depend on the parameters  
$\mymatrix{\lambda},\mymatrix{\eta},\mymatrix{C}$.   
The Fermi-gas ground state of~(\ref{Teffdiag}) is given as   
\begin{equation}  
|\Psi_0  
\left(\mymatrix{\lambda},\mymatrix{\eta},\mymatrix{C}\right)\rangle   
= \prodprime_{\veck,\gamma} \hat{h}_{\veck,\gamma}^+ |\hbox{vacuum}\rangle  
\;.  
\label{psizeroFG}  
\end{equation}
Here, the prime indicates that only those   
single-particle states with   
\begin{equation}
E_{\veck,\gamma}\left(\mymatrix{\lambda},\mymatrix{\eta},\mymatrix{C}\right)  
\leq E_{\rm F}\equiv 0
\end{equation} 
are filled and  
\begin{equation}  
\label{ESP} 
E^{\rm SP}  
\left(\mymatrix{\lambda},\mymatrix{\eta},\mymatrix{C}\right)  
=T_0\left(\mymatrix{\eta}\right)
+E_0\left(  
\mymatrix{\lambda},\mymatrix{\eta},\mymatrix{C}\right)
+ \sumprime_{\veck,\gamma} E_{\veck,\gamma}\left(  
\mymatrix{\lambda},\mymatrix{\eta},\mymatrix{C}\right)  
\end{equation}  
is the energy of the `Fermi-gas' ground state. 
In principle, the variational ground-state energy can now be calculated by a  
minimization of the energy-functional 
\begin{eqnarray}  
E_{\rm c}(\mymatrix{\lambda}, 
\mymatrix{C}  
,\mymatrix{\eta},\Lambda)  
&=&  
E^{\rm SP}  
\left(\mymatrix{\lambda},\mymatrix{C}  
,\mymatrix{\eta}\right)  
+ E^{\rm at} \left(\mymatrix{\lambda},\mymatrix{C} 
\right) -
L\sum_{\sigma\sigma'}\sum_{\alpha \alpha'}  
\left(\eta_{\sigma\sigma'}^{\alpha \alpha'} 
C_{\sigma\sigma'}^{\alpha \alpha'}+{\rm c.c.}\right)
 \nonumber \\  
\label{EC} 
&&   
 +L\Lambda\left(\sum_{\sigma}n_{\sigma\sigma}
\left(\mymatrix{\lambda},\mymatrix{C}\right)-n\right)   
\end{eqnarray}  
with respect to all parameters $x_{i}$ in $\mymatrix{\lambda}, 
\mymatrix{C},\mymatrix{\eta},\Lambda$, i.e., 
\begin{equation}  
\label{EC0} 
E_0^{\rm var}=E_{\rm c}(\mymatrix{\lambda}{}^0, 
\mymatrix{C}{}^0,\widetilde{\eta}^0,\Lambda^0)  
\end{equation}  
where 
\begin{equation} 
\label{cond} 
\left. \frac{\partial}{\partial x_i}   
E_{\rm c}\right|_{\{x_i\}=\{x_i^0\}}=0\;. 
\end{equation}  
Minimizing $E_{\rm c}$ in this straightforward way requires
rather time-consuming calculations of expectation values in  
$|\Psi_0\rangle$ for every single variation of our parameters. Such a  
strategy is prohibitive, in particular due to the large number of 
variational parameters $\mymatrix{\lambda}$.  
Therefore, one needs to develop more sophisticated numerical strategies 
when our general theory is to be applied to realistic model system. 

\section{Landau--Gutzwiller quasi-particles}  
\label{qp}\label{LandauGutzwiller}

The Gutzwiller theory,  
as described in the previous section, is a variational 
approach and by itself a method to examine ground-state properties only. 
As shown in Refs.~\cite{thulpaper,Joerg}, it is possible 
to use the approximate Gutzwiller-correlated ground state 
for the calculation of elementary excitations. 
In particular, the `band structure'~$E_{\veck,\gamma}$ derived in the previous 
section gives the variational excitation energies 
of Landau--Gutzwiller quasi-particles. 
 
\subsection{Definition of single-particle excitations}  
\label{subsec:defLGqp} 

Originally, Gutzwiller-correlated wave functions were introduced as
Fermi-liquid ground states for ordinary metals~\cite{Gutzwiller1963}. 
More generally, they are Fermi-liquid ground states 
in the sense of an adiabatic continuity to 
some non-interacting reference system which can be a Fermi gas,
a BCS superconductor, or a band insulator.
For all such cases, quasi-particle and quasi-hole states 
are readily identified. To this end, we define  
quasi-particle and quasi-hole creation operators as  
\begin{eqnarray}  
 \label{eqn:QT-ER}  
 \widehat{e}_{{\mathbf p}, \tau}^{\, +}& := &  
\widehat{P}_{\mathrm G}   
\widehat{h}{}\vphantom{\widehat{h}}_{{\mathbf p},\tau}^{\, +}   
(\widehat{P}_{\mathrm G})^{-1} \;,\\  
 \label{eqn:QT-VER}  
  \widehat{v}_{{\mathbf p}, \tau}^{\vphantom{\, +}}   
&:=&\widehat{P}_{\mathrm G}  
 \widehat{h}{}\vphantom{\widehat{h}}_{{\mathbf p},\tau}^{\vphantom{\, +}}   
(\widehat{P}_{\mathrm G})^{-1} \;. 
\end{eqnarray}  
These operators obey usual Fermi anti-commutation relations, 
\begin{eqnarray} 
[ \widehat{e}_{{\mathbf p}, \tau}^{\, +}, 
\widehat{v}_{{\mathbf p}', \tau'}^{\vphantom{\, +}}]_{+} 
&=& \delta_{{\mathbf p},{\mathbf p}'}\delta_{\tau \tau'} \mbox{}\;, 
\\ 
{}[ \widehat{e}_{{\mathbf p},  \tau }^{+}, 
 \widehat{e}_{{\mathbf p}', \tau'}^{+}]_{+} 
&=&[ \widehat{v}_{{\mathbf p}, \tau }^{}, 
  \widehat{v}_{{\mathbf p}', \tau'}^{}]_{+}=0 \; ,
\end{eqnarray} 
and they create quasi-particles/quasi-holes in the variational 
Gutzwiller ground-state, as can be seen from 
\begin{eqnarray} 
\widehat{e}_{{\mathbf p}, \tau}^{\, +} 
\widehat{v}_{{\mathbf p}, \tau}^{\vphantom{\, +}}|\Psi_{\rm G}\rangle 
&=& \Theta(E_{\rm F}-E_{{\mathbf p},\tau})|\Psi_{\rm G}\rangle \;,\\ 
\widehat{v}_{{\mathbf p}, \tau}^{\vphantom{\, +}} 
\widehat{e}_{{\mathbf p}, \tau}^{\, +}|\Psi_{\rm G}\rangle&=& 
\Theta(E_{{\mathbf p},\tau}-E_{\rm F})|\Psi_{\rm G}\rangle\;. 
\end{eqnarray} 
We define the quasi-particle or quasi-hole excitation energy  
as   
\begin{equation}  
E_{\pm}(\vecp,\tau)\equiv 
\pm\left(E_{\pm}^{\rm var}(\vecp,\tau)-E_0^{\rm var}\right)
-\mu^{\pm}\left(N_{\pm}^{\rm var}(\vecp,\tau)-N \right)   \;, 
\label{qpdefformula}  
\end{equation}  
where the upper and lower sign corresponds to quasi-particles 
and quasi-holes, respectively. 
Here,
 \begin{equation}  
E_{\pm}^{\rm var}(\vecp,\tau)= 
\frac{\langle\Psi_{\rm G,\pm}^{(\vecp,\tau)} 
|\hat{H} |\Psi_{\rm G,\pm}^{(\vecp,\tau)}\rangle} 
{\langle\Psi_{\rm G,\pm}^{(\vecp,\tau)} 
|\Psi_{\rm G,\pm}^{(\vecp,\tau)}\rangle}
\quad , \quad 
 N_{\pm}^{\rm var}(\vecp,\tau)= 
\frac{\langle\Psi_{\rm G,\pm}^{(\vecp,\tau)} 
|\hat{N} |\Psi_{\rm G,\pm}^{(\vecp,\tau)}\rangle} 
{\langle\Psi_{\rm G,\pm}^{(\vecp,\tau)} 
|\Psi_{\rm G,\pm}^{(\vecp,\tau)}\rangle}
\end{equation}   
are the expectation values of the energy and the average particle number
 in the  quasi-particle/quasi-hole states  
\begin{eqnarray}  
|\Psi_{{\rm G},+}^{(\vecp,\tau)}\rangle  
&\equiv& \widehat{e}_{{\mathbf p}, \tau}^{\, +}|\Psi_{\rm G}\rangle= 
 \hat{P}_{\rm G} \widehat{h}_{\vecp,\tau}^+  
| \Psi_0\rangle \equiv \hat{P}_{\rm G} |\Psi_{{\rm 0},+}^{(\vecp,\tau)}\rangle    \;,\\  
 |\Psi_{{\rm G,}-}^{(\vecp,\tau)}\rangle   
&\equiv& \widehat{v}_{{\mathbf p}, \tau}^{\vphantom{\, +}}|\Psi_{\rm G}\rangle= 
 \hat{P}_{\rm G} \widehat{h}_{\vecp,\tau}  
| \Psi_0\rangle\equiv  \hat{P}_{\rm G} |\Psi_{{\rm 0},-}^{(\vecp,\gamma)}\rangle   \;  .  
\end{eqnarray}  
The one-particle state $| \Psi_0\rangle$ is defined according to~(\ref{psizeroFG})  
in terms of the operators $\widehat{h}_{\veck,\gamma}^+$.  
Note that  $| \Psi_0\rangle$ and $\widehat{h}_{\veck,\gamma}^+$  have to be  
used with their optimum values, given by the parameters  
$\mymatrix{\lambda}{}^0,\mymatrix{\eta}{}^0,\mymatrix{C}{}^0,\Lambda^0$. 
The quasi-particle excitation energy (\ref{qpdefformula}) is   
measured from the (variational) chemical potential $\mu^{\pm}$ which  
describes  the energy for adding a particle to the system or subtracting it, 
respectively,  
\begin{equation}  
\mu^{\pm}=\pm (E_0^{\rm var}(N\pm 1)-E_0^{\rm var}(N))\; .  
\end{equation}  
$E_0^{\rm var}(N)$ is  
the (variational) ground-state energy for a system with $N$~electrons.  
For a metallic or superconducting system, 
$\mu^+=\mu^-=\mu$. For an insulating  
system, $\Delta^{\rm var}=\mu^+-\mu^-$ defines the gap in our variational
theory. The 
variational ground-state energy (\ref{EC0}) depends on the particle density  
$n=N/L$ implicitly, due to the $n$-dependence of all parameters $x_{i}^0$ in 
 $\mymatrix{\lambda}{}^0, 
\mymatrix{C}{}^0,\mymatrix{\eta}{}^0,\Lambda{}^0$, and, explicitly, through  
 the term  $-\Lambda n$ in  (\ref{EC}). Therefore, $\mu_{\pm}$ is given by 
\begin{equation}
\mu_{\pm}=\frac{1}{L} 
\frac{d}{dn}E_{\rm c}(\{x_{i}^0(n)\},n)
=\frac{1}{L} 
\sum_{\{x_{i}\}}\left.\frac{\partial E_{\rm c}} 
{\partial x_{i}} 
\right|_{x_{i}=x_{i}^0} 
\frac{dx_{i}^0 }{dn}-\Lambda_{\pm}^0=-\Lambda_{\pm}^0\;, 
\end{equation} 
where we used~(\ref{cond}) and the fact that, in an insulating 
system, both $\mu_{\pm}$ and $\Lambda^0$ are discontinuous as a function 
of~$n$. 

\subsection{Quasi-particle dispersion}

We write the energies $E_0^{\rm var}$ and $E_{\pm}^{\rm var}(\vecp,\gamma)$ in
(\ref{qpdefformula}) as
\begin{eqnarray}
E_0^{\rm var}&=&\frac{1}{2}
\sum_{\sigma \sigma',\gamma \gamma',\alpha \alpha'}
Q_{\sigma \sigma'}^{\alpha \alpha';\gamma \gamma'}(\mymatrix{\lambda}{}^0,
\mymatrix{C}{}^0)\,
K_{\sigma \sigma'}^{\alpha \alpha';\gamma \gamma'}
(\mymatrix{\lambda}{}^0,\mymatrix{C}{}^0,\mymatrix{\eta}{}^0)+
E^{\rm at}(\mymatrix{\lambda}{}^0,\mymatrix{C}{}^0) \; ,
\nonumber \\ 
E_{\pm}^{\rm var}(\vecp,\tau)&=&
\frac{1}{2}\sum_{\sigma \sigma',\gamma \gamma',\alpha \alpha'}
Q_{\sigma \sigma'}^{\alpha \alpha';\gamma \gamma'}(\mymatrix{\lambda}{}^0,
\mymatrix{C}{}^0
+\mymatrix{\Delta}^{\vecp,\tau}_{\pm}{})\,
\sum_{\veck}\epsilon^{0}_{\sigma \sigma'}(\veck)
\langle\Psi_{\rm 0,\pm}^{(\vecp,\tau)}
|\hat{c}^{\alpha}_{(\alpha\veck),\gamma}
\hat{c}^{\alpha'}_{(-\alpha'\veck),\gamma'}
|\Psi_{\rm 0,\pm}^{(\vecp,\tau)} \rangle \nonumber \\
&&+E^{\rm at}(\mymatrix{\lambda}{}^0,\mymatrix{C}{}^0
+\mymatrix{\Delta}^{\vecp,\tau}_{\pm}{})\;,
\label{h1}
\end{eqnarray}
where
\begin{equation}
K_{\sigma \sigma'}^{\alpha \alpha';\gamma \gamma'}
(\mymatrix{\lambda},\mymatrix{C}
,\mymatrix{\eta})
=\sum_{\veck}\epsilon^{0}_{\sigma \sigma'}(\veck)
\langle\hat{c}^{\alpha}_{(\alpha \veck),\gamma}
\hat{c}^{\alpha'}_{-(\alpha' \veck),\gamma'} \rangle_{0}
\label{kmat}
\;. \nonumber
\end{equation}
The tensor
$\widetilde{\Delta}^{\vecp,\tau}_{\pm}$ gives the change of $\widetilde{C}$
due to the creation of a quasi-particle or a quasi-hole,
\begin{equation}
\widetilde{\Delta}^{\alpha \alpha'}_{\pm;\gamma \gamma'}(\vecp,\tau)
=\pm\left(
V_{\gamma \gamma'}^{\alpha \alpha'}(\vecp,\tau)
-V_{\gamma' \gamma}^{\alpha' \alpha}(\vecp,\tau)
\right) \; ,
\end{equation}
where we introduced
\begin{equation}
V_{\gamma \gamma'}^{\alpha \alpha'}(\vecp,\tau)=
\left(u^{(-\alpha)-}_{\gamma \tau}(\vecp)\right)^{*}
u^{\alpha'-}_{\gamma' \tau}(\vecp)^{}.
\end{equation}
The sum over $\veck$ in~(\ref{h1}) is readily evaluated,
\begin{eqnarray}
\sum_{\veck}\epsilon^0_{\sigma \sigma'}(\veck)
\langle\Psi_{\rm 0,\pm}^{(\vecp,\tau)}
|\hat{c}^{\alpha}_{(\alpha \veck),\gamma}
\hat{c}^{\alpha'}_{-(\alpha'\veck),\gamma'}
|\Psi_{\rm 0,\pm}^{(\vecp,\tau)} \rangle
&=&K_{\sigma \sigma'}^{\alpha \alpha';\gamma \gamma'}
(\mymatrix{\lambda}{}^0,\mymatrix{C}{}^0,\mymatrix{\eta}{}^0)
\\
&&\pm\left(
V_{\gamma \gamma'}^{\alpha \alpha'}(\vecp,\tau)
\epsilon^{0}_{\sigma \sigma'}(\vecp)
-V_{\gamma' \gamma}^{\alpha' \alpha}(\vecp,\tau)
\epsilon^{0}_{\sigma' \sigma}(\vecp)
\right)
\,. \nonumber
\end{eqnarray}
To order $(1/L)^0$ the expansion
of $\delta E^{\pm}_{\vecp,\tau}\equiv 
2(E_{\pm}^{\rm var}(\vecp,\tau)-E_0^{\rm var})$ 
with respect
to $\widetilde{\Delta}^{\vecp,\tau}_{\pm}$ yields
 \begin{eqnarray}
\delta E^{\pm}_{\vecp,\tau}&=&
\pm\sum_{\sigma \sigma',\gamma \gamma'}\sum_{\alpha \alpha'}
Q_{\sigma \sigma'}^{\alpha \alpha';\gamma \gamma'}
\left(
V_{\gamma \gamma'}^{\alpha \alpha'}(\vecp,\tau)
\epsilon^{0}_{\sigma \sigma'}(\vecp)
-V_{\gamma' \gamma}^{\alpha' \alpha}(\vecp,\tau)
\epsilon^{0}_{\sigma' \sigma}(\vecp)
\right)
\nonumber \\ \label{h3}
&&+\sum_{\rho \rho',\alpha \alpha'}\left[\sum_{\sigma \sigma',\gamma \gamma'}
K^{\alpha \alpha';\gamma \gamma'}_{\sigma \sigma'}
\left.\frac{\partial Q^{\gamma \gamma'}_{\sigma \sigma'}}
{\partial C^{\alpha \alpha'}_{\rho \rho'}}\right|
_{\mymatrix{C}=\mymatrix{C}{}^0}
+\left.\frac{\partial E^{\rm at} }{\partial
 C^{\alpha \alpha'}_{\rho \rho'}}\right|
_{\mymatrix{C}=\mymatrix{C}{}^0}
\right]\Delta^{\alpha \alpha'}_{\pm;\rho \rho'}(\vecp,\tau)\\   \nonumber
&&+\sum_{\rho \rho',\alpha \alpha'}
\left[\sum_{\sigma \sigma',\gamma \gamma'}
K^{\alpha \alpha';\gamma \gamma'}_{\sigma \sigma'}
\left.\frac{\partial Q^{\gamma \gamma'}_{\sigma \sigma'}}
{\partial \left(C^{\alpha \alpha'}_{\rho \rho'}\right)^{*}}\right|
_{\mymatrix{C}=\mymatrix{C}{}^0}
+\left.\frac{\partial E^{\rm at} }{\partial
\left(C^{\alpha \alpha'}_{\rho \rho'}\right)^{*} }\right|
_{\mymatrix{C}=\mymatrix{C}{}^0}
\right]
\left(\Delta^{\alpha \alpha'}_{\pm;\rho \rho'}(\vecp,\tau)\right)^{*}
\;.
\end{eqnarray}
We use 
\begin{eqnarray}
\pm E_{\vecp,\tau}&=&
\pm\frac{1}{2}\sum_{\sigma \sigma',\gamma \gamma'}\sum_{\alpha \alpha'}
Q_{\sigma \sigma'}^{\alpha \alpha';\gamma \gamma'}
\left(
V_{\gamma \gamma'}^{\alpha \alpha'}(\vecp,\tau)
\epsilon^{0}_{\sigma \sigma'}(\vecp)
-V_{\gamma' \gamma}^{\alpha' \alpha}(\vecp,\tau)
\epsilon^{0}_{\sigma' \sigma}(\vecp)
\right) \nonumber \\ 
&&+\frac{1}{2}\sum_{\gamma \gamma',\alpha \alpha'}
\left[
\left(
\left(\eta^{\alpha \alpha'}_{\gamma \gamma'}\right)^{0}-
\left(\eta^{\alpha' \alpha}_{\gamma' \gamma}\right)^{0}
\right)
\Delta_{\pm;\gamma \gamma'}^{\alpha \alpha'}(\vecp,\tau)+{\rm c.c.}\right]
\;.
\end{eqnarray}
to evaluate the first line in~(\ref{h3}). 
For the second and the third line  we employ
 \begin{equation}\label{141}
\sum_{\sigma \sigma',\gamma \gamma'}
K^{\alpha \alpha';\gamma \gamma'}_{\sigma \sigma'}
\left.
\frac{\partial Q^{\alpha \alpha';\gamma \gamma'}_{\sigma \sigma'}}
{\partial C^{\alpha \alpha'}_{\rho \rho'}}
\right|_{\mymatrix{C}=\mymatrix{C}{}^0}
=
\left \langle \frac{\partial}{\partial C^{\alpha \alpha'}_{\rho \rho'}}
\hat{T}^{\rm eff} \right \rangle_0
= 
\left.
\frac{\partial E^{\rm SP}}
{\partial C^{\alpha \alpha'}_{\rho \rho'}}
\right|_{\mymatrix{C}=\mymatrix{C}{}^0} 
\end{equation}
and the corresponding equation where $C^{\alpha \alpha'}_{\rho \rho'}$
is replaced by $\left( C^{\alpha \alpha'}_{\rho \rho'} \right)^{*}$.
In (\ref{141}) we have used the fact that the dependence of
$K_{\sigma \sigma'}^{\alpha \alpha';\gamma \gamma'}$
on~$\mymatrix{C}$
can be ignored in the derivative of $E^{\rm SP}$
with respect to $\mymatrix{C}$ because
$|\Psi_0 \rangle$ is an exact eigenstate
of $\hat{T}^{\rm eff}$ and $E^{\rm SP}$ is its eigenvalue. 
Using~(\ref{EC}) and~(\ref{cond}) we can rewrite this expression as
\begin{equation}
\left.
\frac{\partial E^{\rm SP}}
{\partial C^{\alpha \alpha'}_{\rho \rho'}}
\right|_{\mymatrix{C}=\mymatrix{C}{}^0}
=
-\left.
\frac{\partial E^{\rm at}}
{\partial C^{\alpha \alpha'}_{\rho \rho'}}
\right|_{\mymatrix{C}=\mymatrix{C}{}^0}
+\frac{1}{2}\left(
\left(\eta^{\alpha \alpha'}_{\rho \rho'}\right)^0
-\left(\eta^{\alpha' \alpha}_{\rho' \rho}\right)^0
\right)-
\delta_{\rho \rho'}\Lambda_{\pm}^0
\sum_{\sigma}
\left.
\frac{\partial n_{\sigma \sigma}}
{\partial C^{\alpha \alpha'}_{\rho \rho'}}
\right|_{\mymatrix{C}=\mymatrix{C}{}^0} \; ,
\end{equation}
where, again, the same equations hold with $\tilde{C}$ and 
$\tilde{\eta}$ replaced by  $\tilde{C}^{*}$ and $\tilde{\eta}^{*}$, 
respectively.
With these results and
\begin{equation}
N^{\rm var}_{\pm}(\vecp,\tau)-N=\sum_{\sigma}
\left[
\left.\frac{\partial n_{\sigma \sigma}}
{\partial C^{\alpha \alpha'}_{\rho \rho'}}\right|
_{\mymatrix{C}=\mymatrix{C}{}^0}
\Delta^{\alpha \alpha'}_{\pm;\rho \rho'}+{\rm c.c.}\right]
\end{equation}
we find 
$E_{\pm}(\vecp,\gamma)=E_{\vecp,\gamma}$
for the quasi-particle and quasi-hole excitation energy. 
This relations shows that the `band structure'~$E_{\vecp,\tau}$
describes meaningful quasi-particle bands.

\section{Outlook}

Our Gutzwiller theory provides a very good description of the
quasi-particle band structure of ferromagnetic nickel.
We obtain the correct exchange splittings and, in particular,
we reproduce the experimental Fermi-surface topology. 
We find the correct (111)-direction of the magnetic easy axis
and the right order of magnitude of the magnetic 
anisotropy. Our theory even reproduces the experimentally observed
change of the Fermi-surface topology when the magnetic moment
is oriented along the (001)~axis.

Our investigations show that SDFT should be improved
along the following lines. First, it needs to incorporate
an orbital dependence of the exchange-correlation potential
in order to reproduce the anisotropy of the exchange splittings.
It may also be used to correct the partial densities 
of the $3d$ and $4sp$~electrons. Second, 
it should allow an {\rm effective\/} spin-orbit coupling.

Thus far, our calculations have focused on the quasi-particle
band structure at zero temperature. Several extensions of the
Gutzwiller theory are possible. 
First, as shown in~\cite{Joerg},
we can also address magnetic excitations. Therefore,
Gutzwiller theory should reproduce the spin-wave spectrum 
of nickel. Second, we can generalize the idea of elementary excitations
to transitions between atomic states which are induced by photoemission.
In this way we should also be able to make contact
with the experimentally observed `$6\, {\rm eV}$-peak'~\cite{Huefner}.

We emphasize that our Gutzwiller theory as presented 
here is not limited to nickel.
We may apply it equally to other transition metals and their compounds,
for example NiO~\cite{Weiser}.
Our theory also covers superconducting pairing which we may apply
to superconductivity in multi-band systems.
The Gutzwiller approach can and will be applied to 
many interesting correlated-electron problems in the future. 


 
\newpage 

\pagestyle{myheadings} 
\markboth{{\rm REFERENCES}}{{\rm REFERENCES}} 
 
\begin{mybibliography}{99} 
 
\bibitem{Wohlfarth} E.P.~Wohlfarth in {\sl 
Handbook of Magnetic Materials}~{\bf 1}, ed.\ by
E.P.\ Wohlfarth (North Holland, Amsterdam, 1980), p.~1.

\bibitem{vanVleck} J.H.~van Vleck, Rev.~Mod.~Phys.~{\bf 25}, 220 (1953).

\bibitem{Gutzwiller1963} M.C.~Gutzwiller, Phys.~Rev.~Lett.~{\bf 10}, 
159 (1963).

\bibitem{Perdew} J.P.~Perdew and Y.~Wang, Phys.~Rev.~B~{\bf 33},
8800 (1986); A.D.~Becke, Phys.~Rev.\ A~{\bf 38}, 3098 (1988).

\bibitem{Volloneband} M.~Ulmke, Eur.~Phys.~J.~B~{\bf 1}, 301 (1998);
J.~Wahle, N.~Bl\"umer, J.~Schlipf, K.~Held, and D.~Vollhardt,
Phys.~Rev.~B~{\bf 58}, 12749 (1998).

\bibitem{Nolting} W.~Nolting, W.~Borgie\l, V.~Dose, and Th.~Fauster,
Phys.~Rev.~B~{\bf 40}, 5015 (1989).

\bibitem{GW1} L.~Hedin, Phys.~Rev.~{\bf 139}, A796 (1965);
F.~Aryasetiawan, Phys.~Rev.~B~{\bf 46}, 13051 (1992).

\bibitem{Andersen} O.K.~Andersen, O.~Jepsen, and D.~Gl\"otzel
in {\sl Highlights of Condensed-Matter Theory}, ed.~by F.~Bassani, F.~Fumi, 
and M.~Tosi (North-Holland, Amsterdam, 1985), p.~59;
W.R.L.~Lambrecht and O.K.~Andersen, Phys.~Rev.~B~{\bf 34}, 2439 (1986); 
O.~K.~Andersen and T.~Saha-Dasgupta, Phys.~Rev.~B~{\bf 62}, R16219 (2000).

\bibitem{DMFT} For an introduction to dynamical mean-field theory and
its applications to real materials, see, for example, 
K.~Held, I.A.~Nekrasov, G.~Keller, V.~Eyert, N.~Bl\"umer, A.K.~McMahan, 
R.T.~Scalettar, T.~Pruschke, V.I.~Anisimov, and D.~Vollhardt
in {\sl Quantum Simulations of Complex Many-Body Systems: From 
Theory to Algorithms}, ed.~by J.~Grotendorst, D.~Marks, 
and A.~Muramatsu (NIC Series Vol.~{\bf 10}, 2002), p.~175.

\bibitem{Zoelfl} M.B.~Z\"olfl, Th.~Pruschke, J.~Keller, A.I.~Poteryaev, 
I.A.~Nekrasov, and V.I.~Anisimov, Phys.\ Rev.\ B~{\bf 61}, 12810 (2000).

\bibitem{Lichtenstein} A.I.~Lichtenstein, M.I.~Katsnelson, and G.~Kotliar,
Phys.~Rev.~Lett.~{\bf 87}, 067205 (2001).

\bibitem{LDApU} V.I.~Anisimov, J.~Zaanen, and O.K.~Andersen,
Phys.~Rev.~B~{\bf 44}, 943 (1991).

\bibitem{KotliarLDApU} I.~Yang, S.Y.~Savrasov, and G.~Kotliar,
Phys.~Rev.~Lett.~{\bf 87}, 216405 (2001).

\bibitem{Xie} Y.~Xie and J.A.~Blackman, Phys.~Rev.~B~{\bf 69}, 172407 (2004).

\bibitem{GW2} S.~Biermann, F.~Aryasetiawan, and A.~Georges,
Phys.~Rev.~Lett.~{\bf 90}, 086402 (2003).

\bibitem{Donath} M.~Donath, Surf.~Sci.~Rep.~{\bf 20}, 251 (1994).

\bibitem{LB19} M.B.~Stearns in {\sl Landolt--B\"ornstein 
New Series Group~III}, 
Vol.~19A, ed.\ by H.P.J.\ Wijn (Springer, Berlin, 1986), p.~24.

\bibitem{LB23} A.~Goldmann, W.~Gudat, and O.~Rader in
{\sl Landolt--B\"ornstein New Series Group~III}, Vol.~23C2, ed.~by 
A.~Goldmann (Springer, Berlin, 1994).
 
\bibitem{Tsui} D.C.~Tsui, Phys.~Rev.~{\bf 164}, 669 (1967).

\bibitem{Mook} H.A.~Mook, Phys.~Rev.~{\bf 148}, 495 (1966).

\bibitem{EHK78} D.E.~Eastman, F.J.~Himpsel, and J.A.~Knapp,
Phys.~Rev.~Lett.~{\bf 40}, 1514 (1978).

\bibitem{EP80} W.~Eberhardt and E.W.~Plummer, 
Phys.~Rev.~B~{\bf 21}, 3245 (1980).

\bibitem{Moruzzi} V.L.~Moruzzi, J.F.~Janak, and A.R.~Williams,
{\sl Calculated Electronic Properties of Metals}
(Pergamon Press, New York, 1978).

\bibitem{4dTM} This appears to be a general problem 
of the underlying DFT calculation. The same problem appears
for $4d$ transition metals of bcc structure: the position
of the purely $5p$-state $N_{2'}$ also appears shifted by about
$0.7\, {\rm eV}$ to lower energies as compared to the DFT results,
see G.W.~Crabtree, D.H.~Dye, D.P.~Karim, D.D.~Koelling, and J.B.~Ketterson, 
Phys.~Rev.~Lett.~{\bf 42}, 390 (1977).

\bibitem{Singh} M.J.~Singh, J.~Callaway, and C.S.~Wang, 
Phys.~Rev.~B~{\bf 14}, 1214 (1976).

\bibitem{Dalderop} G.H.O.~Daalderop, P.J.~Kelly, 
and M.F.H.~Schuursmans, Phys.~Rev.~B~{\bf 41}, 11919 (1990).

\bibitem{Gersdorf} R.~Gersdorf and G.~Aubert, Physica B~{\bf 95}, 135 (1978).

\bibitem{Gersdorf2} R.~Gersdorf, Phys.~Rev.~Lett.~{\bf 40}, 344 (1978).

\bibitem{PRB} J.~B\"unemann, W.~Weber, and  F.~Gebhard,   
Phys.~Rev.~B~{\bf 57}, 6896 (1998).
 
\bibitem{EPL} J.~B\"unemann, F.~Gebhard, T.~Ohm, R.~Umst\"atter, 
S.~Weiser, W.~Weber, R.~Claessen, D.~Ehm, A.~Harasawa, A.~Kakizaki, 
A.~Kimura, G.~Nicolay, S.~Shin, and V.N.~Strocov, 
Europhys.~Lett.~{\bf 61}, 667 (2003).
 
\bibitem{thulpaper} J.~B\"unemann, F.~Gebhard, and R.~Thul,  
Phys.~Rev.~B~{\bf 67}, 075103 (2003).

\bibitem{SlaterKoster} J.C.~Slater and G.F.~Koster, Phys.~Rev.~{\bf 94},
1498 (1954).

\bibitem{Mattheiss} W.~Weber and L.F.~Mattheiss, 
Phys.~Rev.~B~{\bf 25}, 2270 (1982).

\bibitem{Woelfle} T.~Ohm, S.~Weiser, R.~Umst\"atter, W.~Weber, 
J.~B\"unemann, J.~Low Temp.~Phys.~{\bf 126}, 1081 (2002).

\bibitem{Abragam} A.~Abragam and B.~Bleaney, {\sl Electron
  paramagnetic resonance of transition ions}
(Clarendon Press, Oxford, 1970).
 
\bibitem{BrunoIFFJ} P.~Bruno in {\sl Magnetismus von Festk\"orpern und 
Grenzfl\"achen} (24.~IFF-Fe\-rienkurs, Forschungszentrum J\"ulich GmbH, 1993),
p.~24-1; P.~Escudier, Ann.~Phys.~(Paris)~{\bf 9}, 125 (1975).

\bibitem{nextpaper} W.~Weber et al., in preparation.

\bibitem{Sugano} S.~Sugano, Y.~Tanabe, and H.~Kamimura, {\sl Multiplets of
Transition-Metal Ions in Crystals} (Pure and Applied Physics~{\bf 33},
Academic Press, New York, 1970).

\bibitem{Herring} C.~Herring in {\sl Magnetism, Vol.~IV}, 
ed.\ by G.~T.~Rado and H.~Suhl (Academic Press, New York, 1966), p.~1.

\bibitem{Czycholl} I.~Schnell, G.~Czycholl, and R.C.~Albers, 
Phys.~Rev.~B~{\bf 68}, 245102 (2003).

\bibitem{Hybertsen} M.S.~Hybertson, M.~Schl\"uter, and
 N.E.~Christensen, Phys.~Rev.~B~{\bf 39}, 9028 (1989).

\bibitem{Vielsack} G.~Vielsack and W.~Weber, Phys.~Rev.~B~{\bf 54}, 6614
(1996).

\bibitem{Anisimov} V.I.~Anisimov and O.~Gunnarsson, 
Phys.~Rev.~B~{\bf 43}, 7570 (1991).

\bibitem{Rath} J.~Rath and A.J.~Freeman, Phys.~Rev.~B~{\bf 11}, 2109 (1975).

\bibitem{Kaem90} K.-P.~K\"amper, W.~Schmitt, and G.~G\"untherodt,
Phys.~Rev.~B~{\bf 42}, 10696 (1990).

\bibitem{Callaway} C.S.~Wang and J.~Callaway, Phys.~Rev.~B.~{\bf 15}, 
298 (1977); J.~Callaway in {\sl Physics of Transition Metals 1980},
ed.~by P.~Rhodes (Conf.~Ser.\ Notes~{\bf 55}, Inst.~of Physics, Bristol, 
1981), p.~1.

\bibitem{Joerg} J.~B\"unemann, J.~Phys.~Cond.~Matt.~{\bf 13}, 5327 (2001).
 
\bibitem{Huefner} S.~H\"ufner, {\sl Photoelectron spectroscopy},
2nd edition (Solid-State Sciences~{\bf 82}, Sprin\-ger, Berlin, 1995).

\bibitem{Weiser} S.~Weiser, PhD thesis (Universit\"at Dortmund, 2005),
unpublished.

\end{mybibliography} 
\end{document}